\documentclass[1p]{elsarticle}

\usepackage{hyperref}
\usepackage{amsmath,amssymb, amsfonts,empheq,mathtools,mathrsfs,nccmath}
\usepackage{float,subfigure, wrapfig}  
\usepackage{arydshln,multirow, multicol, rotating, longtable,lscape,booktabs}  
\usepackage{relsize} 
\usepackage{color,xcolor} 
\usepackage{geometry} 
\usepackage[linesnumbered,boxed,ruled,commentsnumbered]{algorithm2e}
\usepackage{tikz} 
\usetikzlibrary{shapes.geometric, arrows} 
\usepackage{verbatim} 
\tikzset{
  treenode/.style = {shape=rectangle, rounded corners,
                     draw, align=center,
                     top color=white, bottom color=blue!20},
  root/.style     = {treenode, font=\Large, bottom color=red!30},
  env/.style      = {treenode, font=\ttfamily\normalsize},
  dummy/.style    = {circle,draw}
} 

\biboptions{authoryear}

\begin{document}
	
	\begin{frontmatter}
		
		\title{Are Crises Predictable? An Review of the Early Warning Systems\\ in the Currency and Stock Markets}
		
		\author[mymainaddress,mysecondaryaddress]{Peiwan Wang}
		\author[mymainaddress,mysecondaryaddress]{Lu Zong\corref{mycorrespondingauthor}}
		\cortext[mycorrespondingauthor]{Corresponding author}
		\ead{Lu.Zong@xjtlu.edu.cn}
		
		\address[mymainaddress]{Department of Mathematical Sciences, Xi'an Jiaotong-Liverpool University.}
		\address[mysecondaryaddress]{111 Ren'ai Road, Dushu Lake Science and Education Innovation District, Suzhou Industrial Park, Suzhou, Jiangsu Province, P.R. China, 215123.}
		
		\begin{abstract}
			The study efforts to explore and extend the crisis predictability by synthetically reviewing and comparing a full mixture of early warning models into two constitutions: crisis identifications and predictive models. Given empirical results on Chinese currency and stock markets, three-strata findings are concluded as (i) the SWARCH model conditional on an elastic thresholding methodology can most accurately classify crisis observations and greatly contribute to boosting the predicting precision, (ii) stylized machine learning models are preferred given higher precision in predicting and greater benefit in practicing, (iii) leading factors sign the crisis in a diversified way for different types of markets and varied prediction periods. 
		\end{abstract}

	\end{frontmatter}

	\section{Introduction}\label{sec:intro}

    \subsection{Crisis predictability}
    The predictability of financial crises is a continually debating open question and yet settled through trans-centuries' arguments among economists and practitioners. The core of controversies focuses on whether the crisis is triggered by unexpected exogenous factors or the market endogenous instability. According to the classic Efficient Market Hypothesis (EMH) theory that is proposed by \citeauthor{Fama1970} in 70's of last century, the securities prices in an ideal market can fully reflect all available information and the financial crises occur as long as external shocks, known as the `black swan' factors, arise. In that situation, governors can contribute little to minimize the economic losses after the crisis. \cite{Fama1970}' theory shows strong assumption on human behaviours' rationality, however market participants are hardly to be always rational and quickly respond to market information in practice. \cite{Arthur107} then relax the strong assumption by pointing out the evolution process of the economy is dynamic, nonlinear and uncertain, which also inherently coincides to \cite{Shciller2000}'s revolutionary proposal of the market endogenous imperfection being mainly induced by the irrational factors. In fact, the essence of this argument fertilizes the flourishing studies on financial crisis forecasting system development as the theoretic basis renders inclination to dilute the assertion that crisis is randomly unpredictable. \cite{Sornette2009} formally advocates the argument that crises are predictable, as the study announced, the crisis is the `dragon-king', not the unpredictable `black swan', since before the crisis abruptly bursts, precursory symptoms like substantial outliers, will be teeming observed in reality. It is inspiring to acknowledge the crisis predictability with a positive attitude given a series of rigorous statistical test on predicted results, even though yet deliver 100\% trust. EMH \citep{Fama1970} and `dragon-king' \citep{Arthur107,Shciller2000,Sornette2009} holders, they both admit the fact that perfectly grasping the exact time of crises seems less attainable, but gaining the awareness on precursory signals of potential financial shocking in advance is achievable.

    \subsection{Comparative early warning system}
    The disputes between two against arguing mainstreams are yet settled, particularly in the contemporary data science background, academics and practitioners have ignited a great volume of diligent works on developing early warning systems to back up the precursory evidence before crisis shocking occurs. The first generation of crisis prediction studies started up in the late 1990's, which concentrates to apply the economic models and statistical approaches \citep{eichengreen1996, Frankel1996Currency, Patel1998,BERG1999,klr1999banking,KLR} to forecast nationwide banking and currency crises. The vulnerability of crisis may cannot be convincingly revealed by first generation of models cluster, as IMF published paper \citep{Berg1999predictable} summarized and exercised on the 1997 Asian currency crisis, their predicting power is barely satisfactory unless under plausible modifications. Machine learning techniques \citep{nag,oh2006,celik2007banking} are then locked to construct the second generation of early warning systems considering its specialization on predicting big size non-linear data. The impetus brought by these state-of-art models accelerates the progress of predicting financial crises in the data driven situation \citep{MarkusH2017MLcomp,BEUTEL2019}, albeit gives concessions to the interpretable expressiveness on detecting leading crisis factors. In general, there are three layers of absences in both first and second generations of crisis predicting model. First, most studies identify the crisis observations on the single index with an arbitrary cutoff, that lacks the elasticity as the back-forward looking historic sample horizons are varied. Second, the good-of-fitness evaluation is either deficient or ambiguous on marking the leading time, such as \cite{BUSSIERE2006,Sevim2014} estimate the probability of crisis occurring within a future horizon interval to soothe the pain of matching to exact crashing time point (this is actually the case of omission for measuring leading time), \cite{KLR, dawood2017predicting} state each leading factors that contribute to the crisis warning thus the leading period is given by counting the periods of leading indicators in advance of the first crisis signal occurs (there are a series of factors leading time), and \cite{KIM2004,oh2006} directly chronicles the crashing event and the produced signaling levels to visualize how the output precedes to the true crises (impacting but not applicable to high-frequency and long-horizon data). Last, the developed predicting system is rarely assessed in practice. Specifically, it is indispensable to test whether the investors' wealth will be gained as the crisis forecasting system participates.

    A board scope of theoretical and experimental research studies on either developing crisis predictive models or investigating leading economic factors for a wide range of countries in various development situations, such as developed ,developing, emerging, and low-income countries \citep{Frankel1996Currency, demirg1998,KOYUNCUGIL2012,BABECKY20131,KIM202094} have been explored by scholars and practitioners. For policymakers, institutional researchers and public investors, to have a competent early warning system (EWS) to forewarn the crisis, is equivalent to grasp the opportunity to counter the risks that potentially lead destructive damaging as well as to make the economy vigorously circulating and the society comfortably steady-going. Thus, how to distinguish robust EWS model among a great quantity of qualified models based on prominent predictive models seems more than momentous. In fact, the horse race among EWS models \citep{MarkusH2017MLcomp} that predict either macroscopic economic crises or market-specific turbulence, has never been suspended, especially in comparing the merits and shortages between the classic regression models and stylized machine learning techniques. \cite{Sevim2014} compare the EWS models developed with binary logistic regressions and machine learning techniques of artificial neural networks (ANN) and decision trees to predict the Turkish currency crisis and accredit that both decision supporting models and ANNs superior to the traditional regression models. The against conservative opinion, as \cite{BEUTEL2019} which systematically examine the out-of-sample performance for almost full series of proposed EWS models hold, deems that the conventional logit regression is fairly efficient to predict systematic banking crises, machine learning based EWS models however need more enhancements before being fully granted in real-world prediction. Such comparison studies, though predicting models are categorized and uniformly verified, yet comprehensively discuss the varied crisis definitions but thetically adopt one classifier for harmonizing comparison metrics in convenience. Therefore, to fully explore the comparative EWS models, various combinations of different types of classifiers and predictive models are required to be implemented and investigated.
    
	\subsection{Objective}
	Our aim is not either to advocate the EWS predicting as mythological oracle or to investigate all proposed EWS models with a overloaded ambition. We effort to preen the pending issues in comparative EWS models with the hope to inspire further exploring sparkles for subsequent researches. To implement this goal, we specify candidate EWS models that will be compared in a unified frameworks with two partitioned constituents to sensibly reason their potentials in addressing crisis identification and warning signal production challenges. In addition to use the chronology critical events, we novelly boostrap the classified samples and calculate the average misspecification rates to reveal the classifier performance in an statistically objective way. To support the investors render better verdicts in a credible and operational way, we propose the EWS participated portfolio constructing process and compare the Sharpe ratios and CER values under varied aversion levels.
	
	The study will be explored in the following sequence. Section \ref{sec:review_ews} will review and summarize the EWS models in terms of crisis classification techniques, predictive models, performance measurements and variable contributing degree estimation. Section \ref{sec:data} shows the empirical results of examining EWS models on Chinese currency and stock markets in terms of classified crisis observations, model predicting robustness and detected leading crisis factors respectively. Section \ref{sec:conclusion} will conclude the compared results and key implications, as well as give further suggestion and discussion.
	
	\section{Methodology Reviews}\label{sec:review_ews}
    Main concerns for developing EWS models embody into two aspects, that respective pros and cons are yet systematically reviewed in terms of structural components of (1) the identification techniques for labelling crisis observations with data frequency variations\footnote{To the best of our knowledge, most studies yet clearly pinpoint how data frequency effect the EWS model performance.}, and (2) the predictive models' forecasting power of producing warning signals in a unified measurement frameworks.
	
    The current developed EWS models are mainly designed for two types of financial crisis: the crisis that concerns the systematic risks, such as the banking crisis and the debt crisis, and the market specific crisis that relates more to iconic price or index dynamics' turbulences, such as the currency and stock market crashes. The classification for first type of crisis gravely relies on central bank's reports, financial institution research publications and experts' opinions \citep{demirg1998, klr1999banking, Lestano2003, beckman, celik2007banking, davis, reinhart2011debt, dawood2017predicting}, which data are either difficult to quantized \citep{davis} or costly to access for the public. Otherwise, the market specific crisis dating technique that commonly labels the crisis level of the market price index or the technically quantized price index (TQI)\footnote{TQI is commonly a composite of weighing several market price index factors. }, such as the exchange market pressure index \citep{frankel, Patel1998, Lestano2003, PENG2008138, yu2010multiscale, Sevim2014} and the market instability index \citep{MII2009, YOON201435, li2015, CHATZIS2018353} are more accessible and processable. 
    
    The TQI classifier is though popular for market crisis studies, generally puzzled by the crisis cutoff determination, in other words, the value of threshold to produce crisis observations is either crudely imposed a arbitrary value (or a fixed percentile) or mildly taken the mean plus a factor of standard deviations, which fails to dynamically adapt to the crisis severity level in different market turbulence scenarios. An alternative path to classify the market specific crisis is in virtue of the Markovian switching regime model, which first clusters observations into several leveled (for example, low- and high-) volatility states and then recognizes crises by taking (conventionally) the top half of the filtering probability for high-volatility state \citep{hamil, hamlin, Abiad2007}. The gap of cutoff optimization is further bridged according to the market turmoil level variation by imposing an automatically thresholding approach on the two-peak methodology theoretical basis \citep{rosenfeld1983Histogram, Ohtsu2007A} on the SWARCH classifier. This dynamically adaptive classifier has been successfully verified in predicting stock crashes \citep{intEWS}. The other issue that is rarely mentioned in previous studies for EWS classifier construction is the variation of data frequency. The crisis observed on low frequently distributed data, such as monthly, quarterly and annually data, includes the pre-crisis effect into the crisis binary dependent variable definition \citep{BUSSIERE2006, Sevim2014, CANDELON2014}, which is reasonable to instruct the long term macro-economic policy adjustments, but yet suitable to direct short term investment portfolio improvement. While, for daily data, to imitate the timeliness in real trading scenarios, it seems more reasonable to eliminate the pre-crisis effect from the crisis variable definition. Except the issue of determining `true' crisis labels, how to validate the classifier's credibility is alike pended, specifically, the way of listing the chronology of historic crisis events or significant turning point for crises against the classifiers' dated crisis episodes \citep{oh2006, Abiad2007} lacks of statistical persuasion. In the study, a more logical validation approach will be proposed to examine the classifier's performance besides the chronological table list.
	
	The predictive models that support EWS constructions will be clustered as three main branches: the logit/probit regression model - which classic methodology pioneers the EWS construction and keeps most prevailing in the crisis prediction studies \citep{eichengreen1996, Frankel1996Currency, BERG1999, BUSSIERE2006, CANDELON2014, dawood2017predicting}; the indicator approach - which provides an alternative nonparametric methodology to detect the leading factors and uses the refined factors to construct the crisis indicator \citep{KLR, klr1999banking, Lestano2003, Berg2005Assessing, COUDERT2008, reinhart2011debt, reinhart2013banking, PENG2008138}; the state-of-art machine learning and deep learning models \citep{nag, oh2006, celik2007banking, yu2010multiscale, YOON201435, CHATZIS2018353, BEUTEL2019, intEWS, SAMITAS2020101507} - which are not merely expert in modeling the data with significant non-linearity and non-normality, but barely subject to the data size as well. Comparing to the first two classic models, the stylized machine learning models generally perform best in forecasting ability, but meanwhile suffers the pain of deteriorating performance on out-of-samples as the model structure complexity is gained that leads serious over-fitting effect \citep{BEUTEL2019,MarkusH2017MLcomp} especially for low frequency data prediction. To evaluate the EWS performance in a unified frameworks, metrics will be categorized into three types: first, the calibration of scores, such as QPS (Quadratic Probability Score), Youden J and SAR; second, the hit-ratio or goodness-of-fit based on the statistics of correct predicted signals and produced false alarms; and third, the back-testing and reality check in practice. The leading factors that greatly contribute to producing crisis signals will also be estimated and compared among three models, such as the parametric coefficients in the logistic regression, the noise-to-signal ratio (NSR) value in KLR approach, feature importance in neural networks and classification trees, and attention weight in attention mechanism hybrid recurrent neural networks \citep{bondEWS}. 
	
	\subsection{Crisis classifiers}
		\subsubsection{Technically quantized index (TQI)} \cite{eichengreen1996}, \cite{Frankel1996Currency} and \cite{KLR} almost simultaneously propose the concept of EWS for currency crisis prediction. To quantize the crisis variable into binary case for zero and one, the market pressure index of $EPI$ is defined. 
		Specifically, the formula for $EPI$ are different in \cite{eichengreen1996} and \cite{KLR}, which are denoted as $EPI_{ERW}$ and $EPI_{KLR}$ in the following formulae respectively.
		
		\begin{align}
		    EPI_{ERW} &= \frac{1}{\sigma_{e}} \frac{\Delta e_{t}}{e_{t}}-\frac{1}{\sigma_{r}}(\frac{\Delta rm_{t}}{rm_{t}}-\frac{\Delta rm_{US,t}}{rm_{US,t}})+\frac{1}{\sigma_{i}}\Delta(i_{t}-i_{US,t})\\
		    EPI_{KLR}&=\frac{\Delta e_{t}}{e_{t}}-\frac{\sigma_{e}}{\sigma_{r}}\frac{\Delta r_{t}}{r_{t}}+\frac{\sigma_{e}}{\sigma_{i}}\Delta u_{t},
		\end{align}
		
		$e_{t}$ denotes the currency exchange rate per US dollars and $\frac{\Delta e_{t}}{e_{t}}$ is the relative change in the exchange rate. $\sigma_{e}$ is the standard deviation for $\frac{\Delta e_{t}}{e_{t}}$. Since the US is the reference country for $EPI_{ERW}$ construction, $(\frac{\Delta rm_{t}}{rm_{t}}-\frac{\Delta rm_{US,t}}{rm_{US,t}})$ and $\sigma_{r}$ are the difference between the relative change in the ratio of foreign reserves and M1 and the reference country and its standard deviation respectively. $\sigma_{i}$ is the standard deviation of nominal interest rate differential of $\Delta(i_{t}-i_{US,t})$. In the $EPI_{KLR}$, $e_{t}$, $\Delta e_{t}$ and $\sigma_{e}$ keep same as $EPI_{ERW}$. $\sigma_{r}$ denotes the standard deviation of the relative change in the gross foreign reserves $\frac{\Delta r_{t}}{r_{t}}$ and $\sigma_{i}$ denotes the standard deviation for the nominal interest rate change $\Delta i_{t}$.
		
		\cite{Lestano2003} and \cite{Lestano2007Dating} then further compare the sensitivity of dating currency crisis under between $EPI_{ERW}$ \citep{eichengreen1996} and $EPI_{KLR}$ \citep{KLR} to different thresholds by varying the value of $\lambda$, and find the $EPI_{KLR}$ is the preferred index with higher sensitivity. 		
		\cite{Sevim2014} validates the conclusion and further proposes a more concise index of financial pressure index ($FPI$) to measure the currency crisis in a standardized average among gross foreign exchange reserves, exchange rate and interest rate, which reasonably synthesizes domestic currency and expels the dispute of choosing the reference country.
		
		\begin{align}
		    	FPI_{t} = \frac{1}{3}(\frac{e_{t}-\mu_{e}}{\sigma_{e}}-\frac{r_{t}-\mu_{r}}{\sigma_{r}}+\frac{i_{t}-\mu_{i}}{\sigma_{i}}),
		\end{align}
	
		where $e_{t}, r_{t}$ and $i_{t}$ are the percentage changes in the exchange rate, monthly gross foreign exchange reserves and monthly change of overnight interest rates at time $t$, respectively. $\mu's$ and $\sigma's$ are the mean and standard deviation value for each accounted terms.
		
		The crisis variable, in our study, thus will be constructed by hiring the binary function of 
		
		\begin{align}
		    C_{t} = \begin{cases}
            1, & \text{if } FPI_{t} > \mu_{FPI}+\lambda_{fpi}\sigma_{FPI}\\
            0, & \text{otherwise.}
        \end{cases} 
		\end{align}
		$\mu_{FPI}$ and $\sigma_{FPI}$ are the mean and standard deviation for the FPI, and $\lambda$ is the coefficient to control the bound to classify crisis observations.

		One of the top cited study of \cite{BUSSIERE2006}, greatly contributes to specify the crisis variable including either the pre-crisis effect or both of the pre-crisis and post-crisis effect, in other words, the binary forward-looking or the ternary forward- and backward-looking variable will be defined to grasp the crisis information in a fixed window of future and past period to rationalize the goal of predicting `whether a crisis occurs within a specific time horizon' and fixing `post-crisis bias that fails to distinguish between tranquil periods and recovery periods'. For the long-term predicting system, the post-crisis bias effect, however, keeps controversial in EWS studies since the crisis variable that covers the sustainable impact distracts the goal of early notifying warning signals. In the study, the binary case which is same defined as \cite{BUSSIERE2006} will be adopted for allowing the general comparisons with other binary classified models \citep{BERG1999,Lestano2003,Abiad2007,davis,Sevim2014}. The definition is listed as follows.
		
		\begin{align}
		    Y_{t} = \begin{cases}
		    1 & \text{if } \exists k=0,...,12\footnote{For long-term prediction, the crisis variable is required to include the pre-crisis effect, thus the `perfect signal' variable \citep{Sevim2014} is referred to take $12$ months as the length of time horizon.} \text{ s.t. } C_{t+k}=1,\\
		    0 & \text{otherwise.}
		    \end{cases}
		\end{align}\label{eq:perfect_signal}

		For stocks, the crisis variable refers to the quantified index of CMAX, which is first proposed in \cite{Patel1998} and further developed in studies of either dating equity market risks or predicting stock crashes \citep{COUDERT2008, li2015, FAUZI2016, fu2019predicting}. CMAX is a ratio index of maximizing the period up to $t$ within the past rolling window size of $m$\footnote{The rolling window size is customarily equal to 24. In our study, the lengthy window however will result in unexpected information loss, hence shorter length of 12 will be substituted}.
		
		\begin{align}
		    CMAX_{t}=\frac{P_{t}}{max(P_{t},...,P_{t-m})}
		\end{align}
		$P_{t}$ is the price index at time $t$. The binary crisis variable for short-term will be similarly defined as follows and the `perfect signaling' variable for long-term keeps the same with equation (\ref{eq:perfect_signal}).
		
		\begin{align}
		    C_{t} = \begin{cases}
		    1, & \text{ if } CMAX_{t} \leq \mu_{CMAX}-\lambda_{cmx}\sigma_{CMAX}\\
		    0, & \text{ otherwise.}
		    \end{cases}
		\end{align}
		
		As Table \ref{tab:crisis_classifier_summary} lists, the upper panel shows calculating formulae for classic TQI defined crisis variables. In this study, to make the classifier generally comparable, the TQI of $FPI$ and $CMAX$ will not use the ad-hoc thresholding coefficient $\lambda_{fpi}=3$ and $\lambda_{cmx}=2.5$ but gradient search a series of values to define the best performed crisis binary dependent variable for currency and stock markets respectively. 
			\begin{sidewaystable}
		\caption{The table summarizes crisis dating methodologies.}
		\setlength{\tabcolsep}{11pt}
		\small
		\begin{center}
			\begin{tabular}{llllr}
				\hline
				Classifier& Market & Index & \multicolumn{1}{l}{Formula}  & Sourced reference\\
				\hline
				&&&&\\
				\multirow{20}{*}{TQI}&\multirow{15}{*}{currency}&\multirow{15}{*}{$EPI$}&$C_{i,t}^{a}=1$ if $\%\Delta e_{i,t}^{b}>25\%$&\multirow{2}{*}{\cite{Frankel1996Currency}}\\
				& &&\quad \quad \quad and $\%\Delta e_{i,t} > (10\%+\%\Delta e_{i,t-1})$&\\
				&&&&\\
				&& &$C_{i,t}=1$ if $EPI_{i,t}>\mu^{c}_{EPI_{t}}+2\sigma^{c}_{EPI_{t}}$ &  \cite{eichengreen1996}\\
				&&&&\\
				&&&$C_{i,t}=1$ if $EPI_{i,t}>\mu_{EPI_{t}}+3\sigma_{EPI_{t}}$&\cite{KLR}\\
				&&&&\\
				&&&$Y^{d}_{i,t}=1$ if $\exists k=1,...,12$ s.t. $C_{i,t+k}=1$&\multirow{2}{*}{\cite{BUSSIERE2006}}\\
				&&&$Y_{i,t}=2$ if $\exists k=0,...,p$ s.t. $C_{i,t-k}=1$&\\
				&&&&\\
				&& &$C1_{i,t}^{e}=1$ if $KLRm_{i,t}^{f} >\mu_{KLRm_{t}}+2\sigma_{KLRm_{t}}$&\multirow{2}{*}{\cite{CANDELON2014}}\\
				&&&$C6_{i,t}=1$ if $\sum_{j=1}^{6}C1_{i,t+j}>0$&\\
				&&&&\\
				&\multirow{2}{*}{currency}&\multirow{2}{*}{$FPI$}&$K_{t}=1$ if $FPI_{t}>\mu_{FPI_{t}}+3\sigma_{FPI_{t}}$&\multirow{2}{*}{\cite{Sevim2014}}\\
				&&&$PS_{t}=1$ if $\exists k = 1,...,12$ s.t. $K_{t+k}=1$&\\
				&&&&\\
				&\multirow{3}{*}{stock}&\multirow{3}{*}{$CMAX$}&$CC_{t}=1$ if $CMAX_{t}\leq \mu_{CMAX_{t}}-2.5\sigma_{CMAX_{t}}$ & \multirow{3}{*}{\cite{li2015}}\\
				&&&$Y_{t}=1$ if $\exists k=0,...,12$ s.t. $CC_{t+k}=1$&\\
				&&&$Y_{t}=2$ if $\exists k=1,...,11$ s.t. $CC_{t-k}=1$&\\
				&&&&\\
				\hline
				&&&&\\
				\multirow{7}{*}{SWARCH}&\multirow{1}{*}{stock}&\multirow{7}{*}{$FPH^{g}$}&$C_{t} = 1$ if $FPH_{t}>0.5$&\cite{hamil} \\
				&&&&\\
				&&& $C_{t} = 1$ if $FPH_{t}>c$,& \multirow{2}{*}{\cite{intEWS}} \\
				&&&where $c$ is the two-peak method optimized cutoff value. &\\
				&&&&\\
				&currency&&$C_{t}=1$ if $1-(1-FPH_{t})^{12}>0.5$,&\cite{Abiad2007} \\
				&&&where $12$ is the length of predictive months.&\\
				&&&&\\
				\hline
			\end{tabular}\label{tab:crisis_classifier_summary}
		\end{center}
		
		\footnotesize{$a$ $C_{i,t}$ is the crisis variable for country $i$ at time $t$.}\\
		\footnotesize{$b$ $\%\Delta e_{i,t}$ denotes the nominal depreciation of the currency for country $i$ in the period $t$.}\\
		\footnotesize{$c$ $\mu$ and $\sigma$ is the mean and standard deviation of the defined index. }\\
		\footnotesize{$d$ $Y_{i,t}$ is the look-forwarding dependent variable with more than two values of $0$ (for non-crises) and $1$ (for crises) being proposed for solving the \textit{post-crisis bias}.}\\
		\footnotesize{$e$ $C1_{i,t}$ is same with $C_{i,t}$ and $C6_{i,t}$ defines the crisis dummy variable conditioning on at least one crisis in the following six months appears, which proposes the forward looking concept to define crisis variable. }\\
		\footnotesize{$f$ $KLRm$ is the modified KLR pressure index, and is formulated as $KLRm_{i,t}=\frac{\Delta e_{i,t}}{e_{i,t}}-\frac{\sigma_{e}}{\sigma_{r}}\frac{\Delta r_{i,t}}{r_{i,t}}+\frac{\sigma_{e}}{\sigma_{i}}\Delta i_{i,t}$. Specific notations in the formula can be found in \cite{CANDELON2014}. }\\
		\footnotesize{$g$ $FPH_{t}$= \textit{filtering probability for high-volatility state} for time $t$.}
	\end{sidewaystable}
	
	\begin{table}
		\caption{The table clarifies crisis dating formulae for currency and stock markets.}
		\setlength{\tabcolsep}{5pt}
		\small
		\begin{center}
			\begin{tabular}{llll}
				\hline
				classifier&market&perfect signal variable&crisis variable\\
				\hline
				&&&\\
				\multirow{5}{*}{TQI}&currency&\multirow{3}{*}{$Y_{t}=1$ if $\exists k=1,..,12$} &\multirow{2}{*}{$C_{t}=1$ if $FPI_{t}>\mu_{FPI_{t}}+\lambda_{fpi}\sigma_{FPI_{t}}$}\\
				&&&\\
				&&s.t. $C_{t+k}=1$ &\\
				&stock&&$C_{t}=1$ if $CMAX_{t}\leq\mu_{CMAX_{t}}-\lambda_{cmx}\sigma_{CMAX_{t}}$\\
				&&&\\
				\hline
				&&&\\
				&currency&&\\
				SWARCH&&$Y_{t}=1$ if $1-(1-FPH_{t})^{12}>0.5$,&$C_{t}=1$ if $FPH_{t}>0.5$\\
				&stock&&\\
				&&&\\
				\hline
				&&&\\
				&currency&&\\
				SWARCH$_{2pk\_opt}$&&$Y_{t}=1$ if $1-(1-FPH_{t})^{12}>c^{opt}$,&$C_{t}=1$ if $FPH_{t}>c^{opt}$\\
				&stock&&\\
				&&&\\
				\hline
			\end{tabular}\label{tab:crisis_classifier_formulae}
		\end{center}
	\end{table}
		
		\subsubsection{Markov regime switching model (SWARCH)}  An flexible alternative to define the crisis by distinguishing the volatility state between tranquil and turmoil periods is inspired by \cite{hamil}, which study proposes the SWARCH model to project the market turmoil level in the business cycle by first clustering the (price) index into different regimes/states and then inferring the probability of observations staying in high-volatility state. Its robustness in financial crisis and contagion detections have been further studied and verified \citep{susram, susedw, regimecopula2018, authorsContagion}. IMF has \citep{Abiad2007} straightforward adopted the SWARCH to predict the currency crisis in Asian countries, which intrinsically defines the crisis dependent variable based on half cutting the filtering probability for high-volatility state (that will be referred as $FPH$ in the following content). Given consistency of conserving the pre-crisis effect for long-term prediction, \cite{Abiad2007} similarly defines a `perfect signal variable' to indulge a $12$ months window length for long-term forecasting. Table \ref{tab:crisis_classifier_formulae} shows the long-term dependent variable $Y_{t}$ based SWARCH as an equivalent transformation of the short-term forecasting\footnote{The deriving process of the transformed equivalence is referred to \cite{Abiad2007}. It constructs on the imposed assumption of the crisis probability in the future months will be neither worsen or improved.}. 
		
		In the study, we adopted AR(p)-SWARCH(K,q) model to classify the stock and currency price index volatility. 
         \begin{align}
       y_{t} &= u + \theta_{1}y_{t-1} + \theta_{2}y_{t-2} +\dots+\theta_{p}y_{t-p}+\epsilon_{t}, \epsilon_{t}|\mathcal{I}_{t-1} \sim N(0, h_{t});\\
        \frac{h_{t}^{2}}{\gamma_{s_{t}}} &= \alpha_{0} + \alpha_{1}\frac{\epsilon_{t-1}^{2}}{\gamma_{s_{t-1}}}+\dots+\alpha_{q}\frac{\epsilon_{t-q}^{2}}{\gamma_{s_{t-q}}}, s_{t} = \{1,\dots, K\}.\label{eq1}
         \end{align}
	$\epsilon_t$ is normally distributed error term with variance of $h_{t}$. $\alpha's$ are non-negative coefficients for modeling scaled variance term, $\gamma's$ are scaling parameters relating to the state-dependent variable $s_{t}$. There are $K$ regimes for $s_{t}$ to present the index volatility states, and the count of regimes is determined by the value of RCM \citep{rcm,bondEWS}, which metric is proven effective to decide the most suitable value for distinguishing the volatility state. 
	
	The filtering probability for high-volatility state (FPH) that is inferred from historic observations $Y_{t}$ can be written as follows
	\begin{align}\label{eq:filt}
	P(s_{t}=\text{high-vol.}|Y_{t};\boldsymbol{\theta}_{t}),
	\end{align}
	$\boldsymbol{\theta}_{t}$ is the parameter vector to be estimated. The crisis variable is henceforth defined as 
	
    \begin{align}
    C_{t} = 
    \begin{cases}
     1, & P(s_{t}=\text{high-vol.}|Y_{t};\boldsymbol{\theta}_{t}) \geq 0.5 \\
    0, & \text{otherwise.}
    \end{cases} 
    \end{align}	
    
    For long-term horizon prediction, we follow the definition in \cite{Abiad2007} as
    
     \begin{align}
    Y_{t} = 
    \begin{cases}
     1, & 1-(1-P(s_{t}=\text{high-vol.}|Y_{t};\boldsymbol{\theta}_{t}))^{12} \geq 0.5 \\
    0, & \text{otherwise.}
    \end{cases} 
    \end{align}	
    
		\subsubsection{SWARCH with two-peak optimized cutoff }
		The arbitrary cutoff in SWARCH classifier has been further improved with greater robustness in \cite{intEWS}, that achieves dynamically identifying stock market turbulences by recursively applying the two-peak method \citep{rosenfeld1983Histogram, Ohtsu2007A}  to each newly included segment of observations to detect the valley bottom point as the optimal cutoff value in the forward moving estimation process\footnote{The algorithm has been specified in \cite{intEWS}.}. In our study, to justify the newly proposed SWARCH classifier on the two-peak method thresholding basis performance on currency market, the study will implement both SWARCH classifiers thereafter. 
		
\begin{figure}[h]
		\centering
		\vspace{-3mm}
		\includegraphics[width=0.5\textwidth]{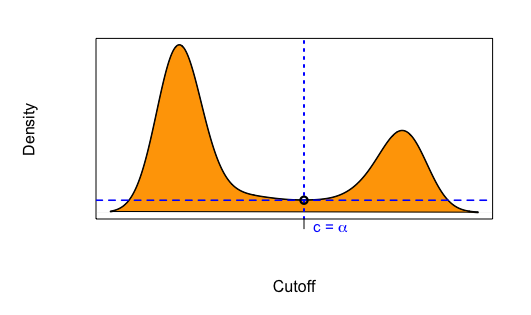}
		\caption{Illustrating the two-peak thresholding methodology: density histogram plot and the concavity detection for optimizing the cutoff value.}
		\label{fig:2pk_struct}
	\end{figure}

		The two-peak method is first proposed and applied to distinguish the difference  between the object and the background grey pixels \citep{Prewitt1966} by locating the frequency density histogram cancavity \citep{rosenfeld1983Histogram, Weszka1978}. The optimizing process of the cutoff value for producing crises is thus inspired by recursively applying the two-peak method to the SWARCH filtering probabilities for the high-volatility state i.e. $P(s_{t}=\text{high-vol.}|Y_{t};\boldsymbol{\hat{\theta}}_{t})$ as expanding the time horizon from $t=l$ to $t=T$, where $l$ is the fixed window size and $T$ is the full sample size. As the example Figure \ref{fig:2pk_struct} shows, the orange density histogram is plotted for filtering probabilities of high-volatility state, then the concavity of histogram is detected at $c=\alpha$, thus $\alpha$ is the optimized cutoff for this time slice. A series of cutoff values will be produced as the time slice observations move forward, that means, the crises will be adaptive to a new cutoff as the volatility information dynamically changes.
	
	The binary crisis variable is thus similarly defined as  \begin{align}
    C_{t} = 
    \begin{cases}
     1, & P(s_{t}=\text{high-vol.}|Y_{t};\boldsymbol{\theta}_{t}) \geq c^{opt}_{t} \\
    0, & \text{otherwise.}
    \end{cases} 
    \end{align}	
    
    For long-term horizon prediction, the crisis formula is written as
    
     \begin{align}
    Y_{t} = 
    \begin{cases}
     1, & 1-(1-P(s_{t}=\text{high-vol.}|Y_{t};\boldsymbol{\theta}_{t}))^{12} \geq c^{opt}_{t} \\
    0, & \text{otherwise.}
    \end{cases} 
    \end{align}	
	where $c^{opt}_{t}$ is the two-peak optimized cutoff value for time $t$.

	\subsection{Predictive models}
	
		\subsubsection{Logistic regression (LR)} 
		Logistic regression is one of most toiling econometric models that is empirically used to construct EWS for predicting curency crisis\citep{eichengreen1996, Frankel1996Currency, BUSSIERE2006}, debt crisis \citep{dawood2017predicting} and financial crisis based on market index and option \citep{li2015}. The advantage of logit regression model sticks two benefits: the latent assupmtion that the dependent variable is linearly linked to other explanatory variables by adding a logistically distributed error, can distinctly convey the relationship among variables and explain the model uncertainty; on the other side, coefficients (with p-value of t-test) magnify the model interpretability and reliability in discovering influential factors. To mitigating the curse of dimensionality in regressing large explanatory variables on one dependent variable, we adopt the stepwise regression to extract and retain the effective combination of variables that maximally explain the dependent variable variation. As mentioned in \cite{BEUTEL2019}, the fixed effect will be removed from the regression since it should be more comparable to other predictive models without extra terms. 
		The Logit regression for modeling the probability of binary crisis variable $y_{t}$ at time $t\in\{1,...,T\}$ can be formulated as follows,
    \begin{align}
    	Pr(y_{t}=1) = \frac{e^{\boldsymbol{x}_{t}\boldsymbol{\beta}}}{1+e^{\boldsymbol{x}_{t}\boldsymbol{\beta}}},
    \end{align}
	where the $\boldsymbol{x}_{t}$ is the vector of explanatory variables at time $t$, $\boldsymbol{\beta}$ is the vector of coefficients. Coefficients will be obtained by maximum likelihood estimation and the joint log likelihood function is written as 
	\begin{align}
		logL = \sum_{t=1}^{T}(y_{t}log(Pr(y_{t}=1))+(1-y_{t})log(1-Pr(y_{t}=1))).
	\end{align}
	In the stepwise backward algorithm, assumed $m$ is the dimension of parameter vector, following steps will be attempted to search for the optimal model.
   \begin{itemize}
   	\item[1.] Establish the regression model between $y$ and all explanatory variables of $\boldsymbol{x}=\{x_{1},x_{2},...,x_{m}\}$, and do $F-$test for each $x$, take the minimum as $F_{l_{1}}=min\{F_{1},F_{2},...,F_{m}\}$.
   	\item[2.] If $F_{l_{1}}>F_{\alpha}(1, T-m+1)$, no variable will be eliminated, the regression model is the optimal one. Otherwise, we elminate $x_{l_{1}}$ and denote the rest of variables as $\boldsymbol{x}_{-l_{1}} = \{x^{1}_{1},x^{1}_{2},...,x^{1}_{m-1}\}$.
   	\item[3.] Establish the regression model between $y$ and $\boldsymbol{x}_{-l_{1}}$, again do the $F-$test for each $x$ and take the minimum as $F_{l_{2}}=min\{F^{1}_{1},F^{1}_{2},...,F^{1}_{m-1}\}$.
   	\item[4.] If $F_{l_{2}}>F_{\alpha}(1, (T-m+1)-1)$, no variable will be eliminated. Otherwise, we elminate $x_{l_{2}}$ and repeat the steps of $F-$test, comparing minimum with the margin and elimination, till not further variable is eliminated from the regression.
   \end{itemize}

     In general, the backward stepwise first put all variables into the model, and then attempt to remove one variable to examine whether significant change appears after the elimination. If there is no significant change, this elimination will be retained until all factors that lead significant change to the model are left. Thus, explanatory variables will be eliminated in turn and finally reordered according to their contribution degree to the model from small to large.
     
		\subsubsection{KLR signal extraction}
		KLR indicator approach is introduced in \cite{KLR} on the basis of nonparametric methodology, which also stands out in the EWS developing realm as this signal extraction approach not only directly assesses the abnormality of single variable behaviour before or during the crisis period without the linearity assumption constraint, but provides a more comprehensive way to policy makers without training background of econometric and statistical modelling as well \citep{klr1999banking, Lestano2003, davis, PENG2008138}, even though in the EWS model comparison study of \cite{Berg2005Assessing} and \cite{davis}, the improved logit regression\footnote{Both study adopt the multivariate logit regression. } is proven to perform better than the signal extraction approach for predicting currency and banking crises. The approach monitors economic variables in a specified period and detects the ones deviates from the noise-to-signal ratio (NSR) minimized threshold as leading factors. The factors that are detected to anticipate the crisis will be counted into constructing the composite indicator by weighing each variable by their respective inverse of NSR \citep{klr1999banking,davis}. The implementing process of KLR methodology is presented as a flowchart diagram, i.e. Figure \ref{fig:KLR_diag}, to simplify words described steps in a more concise way. 
		\usetikzlibrary{calc}
		\tikzstyle{startstop} = [rectangle, rounded corners, minimum width=3cm, minimum height=0.7cm, text centered, draw=black, fill=red!30]
        \tikzstyle{io} = [trapezium, trapezium left angle=70, trapezium right angle=110, minimum width=1cm, minimum height=0.7cm, text centered, draw=black, fill=blue!30]
        \tikzstyle{process} = [rectangle, minimum width=1cm, minimum height=0.7cm, text centered, draw=black, fill=orange!30]
        \tikzstyle{decision} = [rectangle, rounded corners, minimum width=1cm, minimum height=0.7cm, text centered, draw=black, fill=green!30]
        \tikzstyle{arrow} = [thick,->,>=stealth]
        
		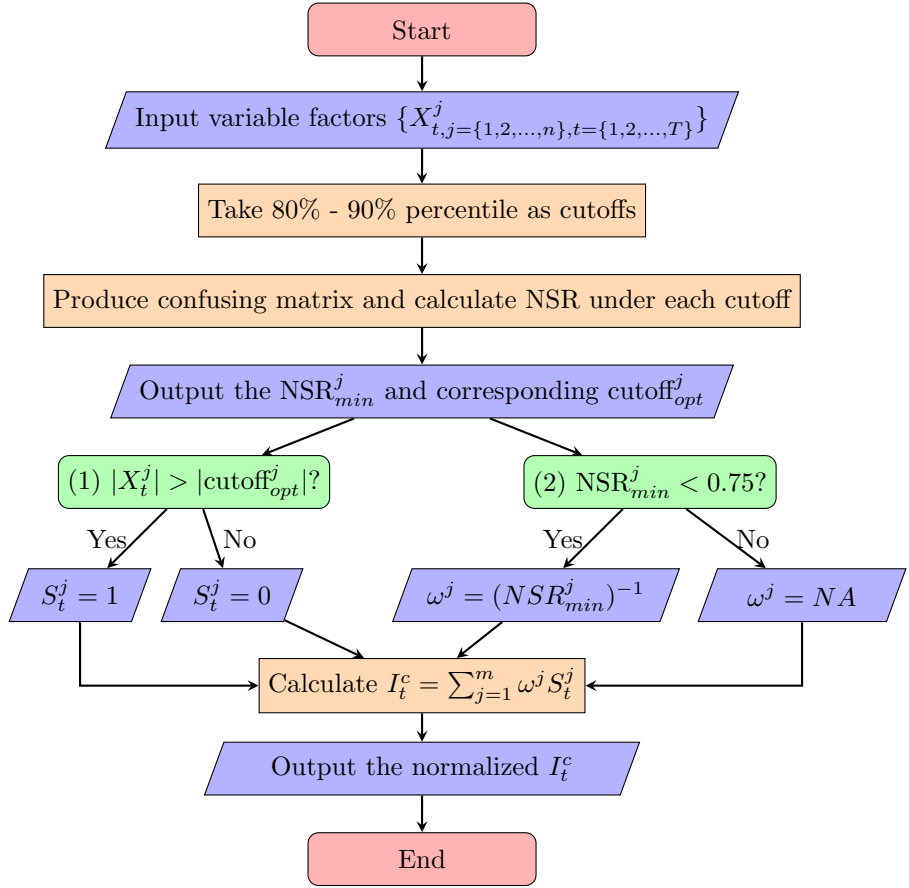
\begin{figure}[!h]
		\begin{center}
		  \begin{tikzpicture}[node distance=1cm]
         \node (start) [startstop] {Start};
         \node (in1) [io, below of=start, yshift= -0.2cm] {Input variable factors \{$X^{j}_{t,j= \{1,2,...,n\},t=\{1,2,...,T\}}$\}};
         \node (pro1) [process, below of= in1, yshift=-0.2cm] {Take 80\% - 90\% percentile as cutoffs};
         \node (pro2) [process, below of=pro1, yshift=-0.2cm] {Produce confusing matrix and calculate NSR under each cutoff};
         \node (out1) [io, below of=pro2, yshift=-0.2cm] {Output the NSR$^{j}_{min}$ and corresponding cutoff$^{j}_{opt}$};
         \node (dec1) [decision, below of=out1, yshift=-0.2cm,xshift=-3cm] {(1) $|X^{j}_{t}|>|\text{cutoff}^{j}_{opt}|$?};
         \node (out2-1) [io, below of=dec1,yshift=-0.5cm,xshift=-1.5cm]{$S^{j}_{t}=1$};
         \node (out2-2) [io, right of=out2-1, xshift=1cm] {$S^{j}_{t}=0$};
         \node (dec2) [decision, right of=dec1, xshift=5cm] {(2) NSR$^{j}_{min} < 0.75?$};
         \node (out3-1) [io, below of=dec2, yshift=-0.5cm, xshift=-1.5cm]{$\omega^{j}=(NSR^{j}_{min})^{-1}$};
         \node (out3-2) [io, right of=out3-1, xshift=2.5cm]{$\omega^{j}=NA$};
         \node (pro3) [process, below of=out2-1, yshift=-0.2cm,xshift=4.5cm]{Calculate $I^{c}_{t}=\sum^{m}_{j=1}\omega^{j} S^{j}_{t}$};
         \node (out4) [io, below of=pro3, yshift=-0.1cm]{Output the normalized $I^{c}_{t}$};
         \node (stop) [startstop, below of=out4, yshift=-0.2cm]{End};
         
         \draw [arrow](start) -- (in1);
         \draw [arrow](in1) -- (pro1);
         \draw [arrow](pro1) -- (pro2);
         \draw [arrow](pro2) -- (out1);
         \draw [arrow](out1) -- (dec1);
         \draw [arrow](dec1) -- node[anchor=east]{Yes}(out2-1);
         \draw [arrow](dec1) -- node[anchor=west]{No}(out2-2);
         \draw [arrow](out2-1) -- ($(out2-1.south) + (0,0)$) node[anchor=east]{}|-(pro3);
         \draw [arrow](out2-2) -- (pro3);
         \draw [arrow](out1) -- (dec2);
         \draw [arrow](dec2)-- node[anchor=east]{Yes}(out3-1);
         \draw [arrow](dec2) -- node[anchor=west]{No}(out3-2);
         \draw [arrow](out3-1) -- (pro3);
         \draw [arrow](out3-2) -- ($(out3-2.south) + (0,0)$) node[anchor=west]{}|-(pro3);
         \draw [arrow](pro3) -- (out4);
         \draw [arrow] (out4) -- (stop);
        \end{tikzpicture}
     \caption{Diagram of implementing the KLR signal extraction approach. NSR$^{j}_{min}$ and cutoff$^{j}_{opt}$ are minimal noise-to-signal ratio and corresponding optimal cutoff for input variable $j$. $I^{c}_{t}$ is the composite crisis indicator \citep{klr1999banking}.}\label{fig:KLR_diag}
		\end{center}
	\end{figure}
	
	\begin{table}[h]
		    \centering
		    \caption{Confusing table for calculating the noise-to-signal ratio for each cutoff.}
		    \setlength{\tabcolsep}{35pt}
		    \begin{tabular}{lcc}
		         \hline
		         & Crisis & No crisis\\
		         \hline
		        Signal was issued & A & B\\
		        No signal was issued & C &D\\
		         \hline
		    \end{tabular}
		    \label{tab:conf_table}
    \end{table}

		In the diagram, we first take 80\% to 90\% percentile of observations for each variable and gradient search the optimal cutoff by producing the confusing matrix, as Table \ref{tab:conf_table} shows, calculating (adjusted) noise-to-signal ratio (NSR) of $\frac{B/(B+D)}{A/(A+C)}$, and searching for the minimal NSR corresponding cutoff value. Then, factor variables will be sifted by the extracted minimal NSR of NSR$^{j}_{min}$ and the optimized cutoff of cutoff$^{j}_{opt}$ for variable $X^{j}$. As green blocks label in the diagram, two decision conditions are (1) whether the variable value is greater than the optimized cutoff and (2) whether the noise-to-signal ratio is smaller than 0.75\footnote{The significant level could be varied for specific markets according to the range of NSR values. Some studies \citep{davis} use 0.5 but find the strict value lead none of factors can be drawn as leading factors.}. Then condition filtered $m$ out of $n$ factor variables will be synthesized by assigning the corresponding inverse of NSR$_{min}$ as their weights to compose the final output, normalized crisis indicator of $I^{c}_{t}$. 

		\subsubsection{Machine learning}
		Machine learning models are the state-of-art techniques that are more flexible than traditional econometric models to predict on complex data with non-linearity. A wide range of EWS models constructed on machine learning techniques, such as neural networks \citep{nag,oh2006,celik2007banking,Dong2009,Sevim2014,YOON201435}, decision trees \citep{TANAKA2016RF,SAMITAS2020101507,MarkusH2017MLcomp}, support vector machine \citep{AHN20112966}, and deep neural networks \citep{intEWS,bondEWS}, have been studied. According to 
       former comparison work \citep{BEUTEL2019,CHATZIS2018353} and our previous studies on stock and bond markets \citep{intEWS,bondEWS}, we select the neural networks, the tree model of random forest and gradient boost and the attention based long-short term memory networks as the candidate machine learning models, to explore deeper comparisons as well as to cover our preceding outstanding work. 
       
\begin{figure}[h]
		\centering
		\includegraphics[width=0.75\textwidth]{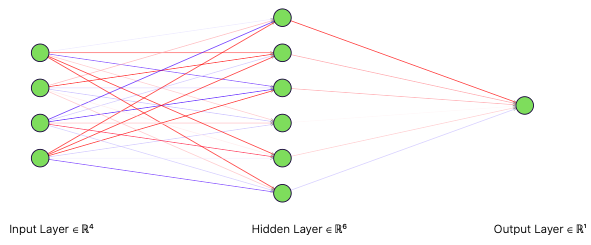}
		\caption{The structure of a sample three-layer artificial neural networks. Green circles are cells of activation functions to process information before passing through the corresponding layer. The arrows represent the information flow direction from input to output, where red and blue label the positive and negative edge proportional to assigned weights.}
		\label{fig:NN_struct}
	\end{figure}

       \textbf{Neural Networks.} 
       As the booming computational vision and big data science unseal a nova technology era, more powerful models are required to solve the nonlinear problems with greater precision and less cost. The artificial neural networks (ANN) are born in this background, and by far, in spite of brimming with disputes on the transparency and interpretability, are generally acknowledged as the most robust and flexible model especially for predicting work. In the ANN applications on EWS model construction, \cite{nag}, \cite{oh2006} and \cite{Fioramanti2008} have successfully predicted the currency, stock and debt crises by hiring the feed-forward multi-layer perceptron. Figure \ref{fig:NN_struct} shows such architecture of a three-layer neural networks embedded on 4-cell input later, 6-cell hidden layer and 1-cell output layer. In practice, the count of neurons in the input layer is required to be same with the dimension of input predictors and the cell number determination for hidden layer will be in trials that start from 2 and increase at a rate of 2 power\footnote{The experimental options for hidden layer structure are 2,4,8,16,32.}. As increasing the structural complexity will influence the networks predicting performance, especially highly winded networks will bring over-parameterization and then make the model less generalized beyond the trained samples, some techniques such as \textit{drop out} and \textit{early stopping} thus will be hired to alleviate the over-fitting problem. 
       
       The neurons in each layer provide driving force to aggregate information by hiring activation functions. A plenty of activation functions, such as $sigmoid$, $ReLU$ and $tanh$, are available to process various nonlinear relationships according to the property of learning target. In the study, we hire $ReLU$ and $sigmoid$ activation functions for the hidden and the output layers respectively. They have formulations as follows.
       
       \begin{align}
          \text{ReLU: }f^{1}(x) & = \begin{cases}
          x, & \text{if } x>0,\\
          0, & otherwise.
          \end{cases}
          \\
           \text{sigmoid: }f^{2}(x) &= \frac{1}{1+e^{-x}}.
        \end{align}       
       Before the output being processed, the weight parameters vector will be applied for each layer neurons, thus the aggregated information can be normally connected. Denote $\boldsymbol{w}^{1}$ and $\boldsymbol{w}^{2}$ to be the weight parameters for bridging between (1) input layer and hidden layer and (2) hidden layer and output layer. Thus, the predicted result for a three-layer ANN with $n$ input variables, $m$-cell hidden layer and single cell output layer, can be written as follows,
       
       \begin{align}
           \hat{y}_{t+1} = f^{2}(\sum^{m}_{j=0}w^{2}_{j}\cdot f^{1}(\sum^{n}_{i=0}w^{1}_{j,i}\cdot x_{i,t})),
       \end{align}
       where $x_{i,t}$ is the value of variable $i$ at time $t$, $w_{j,i}$ is the applied weight to the $i^{th}$ input neuron for producing the input for $j^{th}$ hidden neuron and $w_{j}$ is the applied weight to $j^{th}$ hidden neuron output for the final singular prediction. For more than three layers model, the process can be recursively implemented by assigning various dimensional weight parameters. The parameters will then be optimized by minimizing the $L_{2}$ penalized objective function in 100 epoch iterations.
  
    \textbf{Random Forest and Gradient Boosting Tree.} Both random forest and gradient boosting tree are tree-based model on the ensemble learning base. The core idea for decision tree is to continuously partition data into homogeneous clusters by refining the selection rules as either building or pruning tree branches to get the optimal tree structure. The tree-based model can naturally visualize the catergorizing rules and extract the variable importance, the model interpretability is thus more remarkable than other machine learning techniques. \cite{KOYUNCUGIL2012} and \cite{TANAKA2016RF} use the tree-based model construct EWS for predicting risk pressure for small enterprises and nationwide bank failures, which alters the practitioners' perspective in nonparametric models' predicting power. We thus put two advanced tree-based models in the stylish ensemble learning technique, random forest and gradient boosting tree, into model contrasts. 
    \begin{figure}[H]
     \centering
      \begin{tikzpicture}
  [
    grow                    = right,
    sibling distance        = 6em,
    level distance          = 11em,
    edge from parent/.style = {draw, -latex},
    every node/.style       = {font=\footnotesize},
    sloped
  ]
  \node [root] {Split1}
    child { node [env] {Model6}
      edge from parent node [below] {rule 1-b} }
    child { node [env] {Model1, Split2}
      child { node [env] {Model3, Split3}
        child { node [env] {Model5}
          edge from parent node [above] {rule 3-b} }
        child { node [env] {Model4}
                edge from parent node [above] {rule 3-a} }
        edge from parent node [below] {rule 2-b} }
      child { node [env] {Model2}
              edge from parent node [above, align=center]
                {rule 2-a} }
              edge from parent node [above] {rule 1-a} };
\end{tikzpicture}
        \caption{An example of tree model structure with three clustering rules, three splits and six regression models. Model 4, for instance, will be implemented by all split predictors from 1 to 3 and rule 1-a, 2-b and 3-a filtered data points.}
        \label{fig:sample_tree}
    \end{figure}
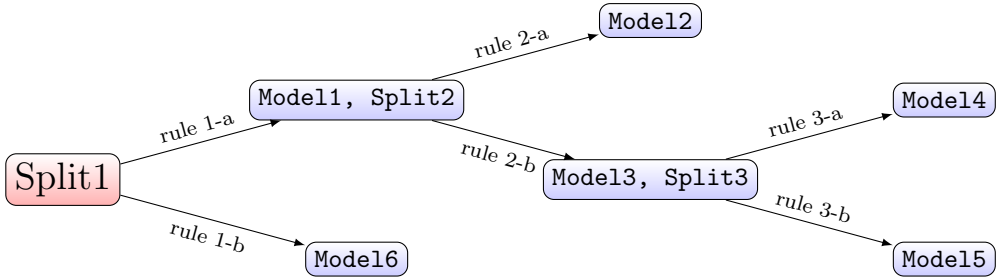
   
    Random forest solves the decision tree's weakness in generalization by growing multiple trees with boostrapping aggregation. The process is implemented as follows.
    
    \begin{itemize}
        \item[1.] Boostrap a sample from original data and build a tree on the boostrapped sample.
        \item[2.] On each split of the trained tree, randomly select $k$ features from original $n$ predictors, where $k\leq n$, then determine the best one among $k$ features and partition data.
        \item[2.] Repeated step 1 and 2 $M$ times, thus $M$ different decision trees are built with corresponding randomly combined $k$ features and partitioned data.
        \item[3.] The best tree model will be determined by monitoring the error value that continuously decrease to most stable level.
        \item[4.] Keep the best performed model and extract the variable importance.
    \end{itemize}
    Hence, from the implementation process for random forests, tuning parameters of $k$ and $M$ are inevitable to designated. The default value for $k$ is equal to $\frac{n}{3}$, one third of predictors count. While, in the study, we alter to run a loop for $k$ taking from $1$ to $(n-1)$ and search for the most efficient value that minimizes the error rate\footnote{The R package `randomForest' implements the random forest model.}. For determination of $M$, the iteration will first run $200$ times til stabilize the error value in a low level band. The value for $M$, according to our experiments, is around $50-80$\footnote{The value of $M$ is fluctuating for different data frequency.}.
    
    Gradient boosting machines make the tree-based model algorithm more adaptive. It shares the similarity of random forest that the final prediction is produced on an ensemble of tree models, but its constructing way is substantially different. Trees in random forests are built independently and each one will reach the maximum depth, while in gradient boosting, the trees are dependent on previous fitted trees by allowing the minimum depth. The computation steps are listed as follows. 
    
    \begin{itemize}
        \item[1.] Initialize $D$ and $M$ to be the tree depth and number of iterations. Compute the average of response $\Bar{y}$ as the initial predicted value.
        \item[2.] Start from the first iteration $1$, calculate the residual, the difference between predicted value and observed value, and fit a $D$ depth tree by setting the residuals as response.
        \item[3.] Produce new predictions by using the fitted tree.
        \item[4,] The predicted value will thus be updated by recursively implementing the step 2 and 3 and adding up the previous predicted value from past iterations.
    \end{itemize}
	
	Similarly, $D$ and $M$ are the tuning parameters for the gradient boosting machine. In the study, we across validate $D=\{1,2,...,5\}$ and $M=\{50,100,150,...,500\}$, find the combination of $D=3$ and $M=100$ performs best by maximizing the AUC value for binary classification\footnote{R package of `xgboost' helps to fit the gradient boosting model.}. The time cost of implementing the gradient boosting machine is more pricey than random forest since the random forest constructs independent trees in parallel, the gradient boosting, though restricts the tree grown depth, aggregates previous results in an adaptive recurring process.
	
\begin{figure}
        \centering
        \includegraphics[width=0.65\textwidth]{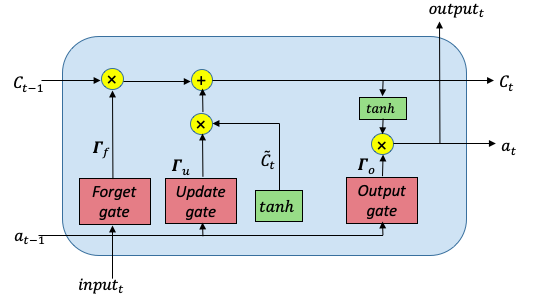}
        \caption{The inner structure of LSTM cell. $\mathbf{\Gamma}_{f}$, $\mathbf{\Gamma}_{u}$, $\mathbf{\Gamma}_{o}$ are sigmoid functions of the forget gate, the update gate and the output gate that determine the information to be discarded, added and reproduced. $\tilde{C}_{t}$ is the new candidate output created by the $tanh$ layer, which is limited in the range $[-1,1]$. $a_{t}$ and $C_{t}$ are recurrently employed activation function and peehole function that carry historic information flow to the future memory block. The initial values of $C_{0}$ and $a_{0}$ are both zero. }
        \label{fig:lstm}
    \end{figure}

	\textbf{Attention based Long-short term memory networks.}
    The deeper variant of neural networks, the multi-layer perceptron models that inherently suit the time dependency, such as recurrent neural networks (RNN) and long-short term memory networks (LSTM), have been generally applied in financial prediction studies \citep{fischer2018deep,liu2019novel,cao2019financial} and present more promising forecasting power than plain neural networks. As Figure \ref{fig:lstm} shows, the input at time $t$ will access the LSTM perceptron and then pass through three `gates' and a series of processing calculations to purify and aggregate information, and in final, carry the activation function $a_{t}$ and peehole function $C_{t}$ to next cell and produce (intermediate output for) predicted results. Specific formulations for the LSTM cell can be written as follows.
    
     \begin{align}
	    \mathbf{\Gamma}_{f} &= \sigma(x_{t}U^{f} + a_{t-1}W^{f}),\\
	    \mathbf{\Gamma}_{u}&=\sigma(x_{t}U^{u} + a_{t-1}W^{u}),\\
    	\mathbf{\Gamma}_{o}&=\sigma(x_{t}U^{o} + a_{t-1}W^{o}),\\
    	\tilde{C}_{t}& = tanh(x_{t}U^{g}+a_{t-1}W^{g}),
      \end{align}
    where $\sigma's$ refer to activation function for each gate, $x_{t}$ is the input at time $t$, $U's$ and $W's$ are assigned weight parameters to connect two neighbour cells by passing processed information from past to future. 
    
\begin{figure}[!h]
        \centering
        \includegraphics[width=0.7\textwidth]{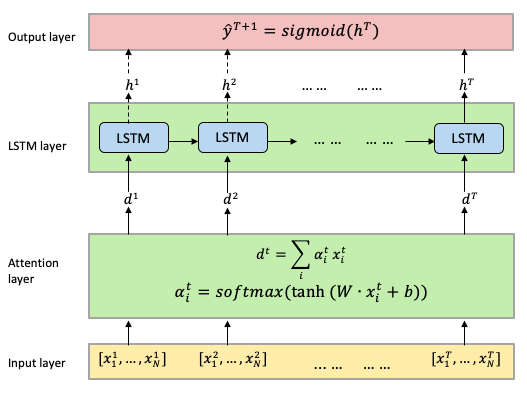}
        \caption{The attention based LSTM networks structure. $T$ is the preset time step size. Vectors of observed variables $[x^{t}_{1},...,x^{t}_{N}]$ at time $t$ for $t=\{1,...,T\}$ will be processed by activation function of $tanh$ and probability transformation function of $softmax$ given uniformly distributed weight vector $W$ and a constant vector $b$. Then, attention filtered information will be passed to the LSTM layer for further processing til final predicted results $\hat{y}^{T+1}$ is generated. The dotted arrows that direct $\{h_{t, t=\{1,...,T-1\}}\}$ means the intermediate output before $T$ will not be memorized by LSTM cell and further be participated into the final prediction.}
        \label{fig:attn_lstm}
    \end{figure}

    The LSTM based EWS has been proven more effective in the comparison to other two machine learning predictive models, back-propagation neural networks (BPNN) and support vector machine (SVM), for predicting stock crashes \citep{intEWS}. The model is then made more comparable in terms of extracting variable importance by stacking a upper attention layer on the LSTM layer \citep{bondEWS}, as Figure \ref{fig:attn_lstm} shows, before entering into LSTM cells, attention mechanism will filter out high-value information among a large amount pieces of inputs. It revolutionizes the weakness of traditional way that assigns same weight vector to each input, by varying the weight according to inputs' value. Thus, a sequence of $\{h^{t}_{t=\{1,...,T\}}\}$ will be made from a network of $T$ LSTM memory cells that process the generated output $\{d^{t}_{t=\{1,...,T\}}\}$ from attention layer, and the final prediction $\hat{y}^{T+1}$ will be produced by a sigmoid function to squash the value into [0,1]. The process then will be recursively carried out on sequential sample pieces from $t=\{2,...,T+1\}$ onward till the end at $\boldsymbol{\mathcal{T}}$.
    
	\subsection{Measurements for classifier and predicting robustness}
	Three types of measurements will be adopted to examine the EWS effectiveness in terms of classifiers, predicted results and practical usefulness.

	\subsubsection{Misspecification rate}\label{sec:misspecification_proc}
	For the crisis classifier, except matching to the chronology critical events, we take an objective way of first calculating the misspecification rate for identified crises on the out-of-sample size truncated full samples and the out-of-samples for each classifier, and recognizing the best performed one with the least misspecification rate. The performance is then further validated by statistically boostrapping the $1000$ sample pieces from full observations and take the average misspecifications.
	
	Table \ref{tab:crisis_classifier_formulae} lists the classifiers that will be put in contrast, thus a justification approach will be proposed to investigate the classifier identifying precision as compare with `true' crisis observations. This judgement is yet objectively committed because the exact timing of `true' crisis is ambiguously personalized corresponding to individuals' cognition level on crises. Most studies alter to match the either critical events or significant signs \citep{oh2006, Abiad2007, Sevim2014, Eijf2019} that once appeared in notorious financial crisis periods to the identified results, but this judgement somewhat lacks of persuasion. The study will though, in the intuitive judgement way, collect the chronologically published evidence to clarify the classifier's identifying performance, a more statistical-orient evaluating method to compare the identified crises difference between for full samples and for test set, as following itemized procedure described, will be proposed and implemented to avoid the subjectivity. 
	
	\begin{itemize}
		\item[1.] Classifiers will first identify crisis observations for full samples (with size of $N_{f}$) and test set (with size of $N_{t}$) respectively, then identified crises on full sample is the vector of $\boldsymbol{C}_{f}=\{C_{1},C_{2},...,C_{N_{f}}\}$ and that on test set is $\boldsymbol{C}_{t}=\{C_{1},C_{2},...,C_{N_{t}}\}$;
		\item[2.] Full samples will be truncated as the same time horizon of test set and then the identified crises on truncated full samples will be denoted as the vector of $\boldsymbol{C}_{f_{truc}}=\{C_{N_{f}-N_{t}+1},...,C_{N_{f}}\}$;
		\item[3.] Misspecification rate on the test set will be calculated as 
		\begin{align}
		\text{misspefication rate} = \frac{\sum^{N_{t}}_{i=1}(\boldsymbol{C}_{f_{truc}}\neq\boldsymbol{C}_{t})}{N_{t}},
		\end{align}
	\end{itemize}
	
    To avoid the opportunist result, boostrapping will be further required to validate the classifier average performance, and the most preferred classifier will give the lowest average misspecification rate on boostrapped samples. The implementing procedure is to repeat the above step 1-2 on the randomly selected piece of samples with the size of $N_{t,j}$ for $B$ times\footnote{$B$ is usually great larger than the sample size.}. Then, from $j=1$ to $j=B$, the average misspecification rate be calculated as

		\begin{align}
		\text{avg. misspecification rate} = \frac{1}{B}\sum_{j=1}^{B}\Bigg[\frac{\sum^{N_{t,j}}_{i=1}(\boldsymbol{C}_{f_{truc},j}\neq\boldsymbol{C}_{t,j})}{N_{t,j}}\Bigg]
		\end{align} 

	\subsubsection{Statistical metrics}
    To examine the predicting robustness, as previous EWS studies adopt, two types of statistical measurements will be taken (1) the goodness-of-fit, such as the ratio of correctly called crisis observations and crisis onsets, and false alarms \citep{BERG1999,BUSSIERE2006,davis,dawood2017predicting} on out-of-samples, and (2) credit-score calibrations, such as quadratic probability score (QPS), Youden index $J$ \citep{CANDELON2012} and SAR\footnote{The composite score of averaging the accuracy, AUC and 1-RMSE.} \citep{sar2004}.
	    
	\begin{align}
	&\text{Quadratic Probability Score}:  QPS = \frac{1}{T}\sum_{t=1}^{T}2(\hat{y}_{t}-y_{t})\\
	&  \text{Youden $J$}: J  = \text{sensitivity}+\text{specificity}-1 \\
	& \text{SAR}: \text{SAR} = \frac{1}{3}(\text{Accuracy} + \text{AUC} + (1- \text{RMSE}))
	\end{align}
    
    \subsubsection{Invigilation in practice}
    
	To invigilate the predicted signals usefulness in practice, back testing and reality check will be hired in assigning varied portfolio weights according to the investors' risk aversion in turmoil periods. The risk aversion level is proven dynamic as the financial situation shifts from tranquility to turmoil, specifically, both macro- and micro-level shocks increase the investors risk aversion level \citep{SAKHA2019184}, and then the asset allocation will be necessarily re-adjusted to improve the portfolio returns \citep{HASLER2019182} and lower the amplifying degree from institutional investors' negative behaviour to exacerbate subsequent crashes \citep{Fan2020}.
	
	The implementing back test process is on the basis of constructing a moving forward the dynamics for asset allocation by referring to the warning signals. Similarly to previous studies for portfolio construction \citep{RAPACH2010,DAI2020101216}, in the study, for each asset of $\{r_{i,i=1,...,S}\}$, at time $t$, the portfolio weight will be given as  
	\begin{align}
	    w_{i,t} = \frac{1}{\gamma+\hat{y}_{i,t+1}}(\frac{\eta_{i,t}}{\sigma_{i,t}}),
	\end{align}
   where $\eta_{i,t}$ is the return rate for asset $r_{i}$ at time $t$, $\sigma_{i,t}$ is the realized volatility with one-year rolling window for asset $r_{i}$ at time $t$. $\gamma$ denotes the initial risk aversion level for investors, which will be varied from $1$ to $3$ to represent the innate risk preference tastes from low to high in the real market. $\hat{y}_{i,t+1}$ is the produced crisis warning signal from each EWS model for asset $r_{i}$. Here, the short sales is restrained no more than 50\% and the long position is allowed no more than 150\%, the value of $w_{i,t}$ will be hence limited into $(-0.5,1.5)$ \citep{Campbell2007,DAI2020101216}.
   
   Sharpe ratios and the average certainty equivalent return (CER) will be then calculated as follows,
   \begin{align}
       & \text{Sharpe ratio} = \frac{\overline{R_{p}}}{\sigma_{p}}\\
       & \text{CER} = \overline{R_{p}}-\frac{1}{2}\gamma\sigma_{p}
   \end{align}
   where $\overline{R_{p}}$ and $\sigma_{p}$ denote the mean and standard deviation of portfolio returns, $R_{p} = \sum^{S}_{i=1}w_{i,t}\eta_{i,t}$.
   
   The reality check will be implemented by revisiting the Eq. \ref{eq:f}, portfolios’ realized variance \citep{intEWS} between the EWS involved and non-EWS benchmarks, and make judgement on whether the null hypothesis of EWS not improving the return variance level, i.e.  $\text{H}_{0}:E(f)\geq 0$, will be rejected after stationary boostrapping 1000 times for p-value calculation.
   
   \begin{align}
       f_{t} = V_{EWS,t}-V_{Bench,t}\label{eq:f}
   \end{align}
  
   \section{Experimental study on currency and stock markets in China}\label{sec:data}
	
   As the world's second largest economy, financial markets in China have been continuously improving in a blossom development and rapidly expanding international economic interactions. It is, however, the high-speed ``Oreint Express'' opts to an unpredictably changing economic situation and follows the endeavour to maintain the market stability, especially for two principal markets of currency and stocks, in the financial crisis triggered turbulence. For the currency market, after the ``8.11'' Exchange Rate Reform in 2005, the Central Bank has been reduced their intervention in manipulating the exchange-rate to lowest portion. According to the released data released from the Bank for International Settlements (BIS), the nominal effective exchange rate of the Chinese yuan (CNY) appreciated 38\% from the beginning of 2005 to June 2019, and the real effective exchange rate appreciated 47\%, which not only leads China to be one of the largest currencies in the globe, but gains the exposure risk to financial crises as well. For the Chinese stock market, it has experienced tremendous ups and downs in financial crises and also has been playing as the barometer of economic situation for more than three decades, from the original grassroots era (1990-1995), the golden age (1995-2000) to the era of returning to reason (2000-2005) and prosperity (2005-2010), and to the most recent fanaticism caused stock crash (2015-2016). The possibility of constructing an effective predictive system to monitor two primary investment markets turbulence will gain the economy reliance to crash risks and guide the policymakers and investors to avoid devastating losses in the withering attacks. 
   
   To fulfill the aim of comparing EWS model predicting effectiveness on market specific crises and unravel mystery of whether any forecasting system will work on Chinese markets, which were once thought to be less transparent and highly government-manipulated, we thus experiment on both currency and stocks markets of China.

	\begin{table}[htpb]
		\caption{Summary of factor variables.}
		\setlength{\tabcolsep}{4pt}
		\small
		\begin{center}
			\begin{tabular}{p{80pt}p{30pt}p{130pt}p{70pt}r}
				\hline
				Sector& 
				Label&
				Variable& 
				Frequency&
				\vspace*{0.1mm}
				Source\\
				\hline
				\multirow{6}{*}{Stock market}&ssec&Shanghai composite index&daily&\multirow{6}{*}{WIND}\\
				&ssec\_r&return rate of Shanghai composite index&daily&\\
				&sez&Shenzhen composite index&daily&\\
				&sez\_r&return rate of Shenzhen composite index&&\\
				\hline
				\multirow{3}{*}{Currency market}&frx&USD/CNY exchange price&daily&\multirow{3}{*}{WIND}\\	
				&frx\_r&return rate of USD/CNY exchange price&daily&\\
				\hline
				\multirow{2}{*}{Interest rate}&dis\_r&discount rate&monthly&St. Louis Fed\\
				&int\_r&demand deposit interest rate&monthly&WIND\\
				\hline
				\multirow{30}{*}{National economy}&bal\_g&balance of international payments (\% in GDP)&annually&\multirow{30}{*}{WIND}\\
				&cci&consumer confidence index&monthly&\\
				&cpi&consumer price index (year-on-year basis)&monthly&\\
				&comd\_g&commodity trade (\% in GDP)&annually&\\
				&deb\_g&total debts (\% in GDP)&annually&\\
				&dom\_c&domestic credit&monthly&\\
				&dom\_g&banking departments offered domestic credit (\% in GDP)&annually&\\
				&fix\_a&completed investment in fixed asset(cumulative year-on-year basis)&monthly&\\
				&frx\_res&official foreign exchange reserve&monthly&\\
				&gdp\_p&per capital GDP&annually&\\
				&gdp\_p2&per capital GDP (year-on-year basis) &annually&\\
				&gdp\_c&GDP growth contribution rate (to consumption)&annually&\\
				&im\_ex&total import-export volume(year-on-year basis)&monthly&\\
				&ind\_g&industrial added value (\% in GDP)&annually&\\
				&ind\_p&production and sales ratio of industrial products&monthy&\\
				&ind\_v&industrial added value (year-on-year basis)&monthly&\\
				&inf\_r&inflation rate&annually&\\
				&mac\_c&macro-economic climate index&monthly&\\
				&m2\_g&board money supply (\% in GDP)&annually&\\
				&ppi&producer price index (year-on-year basis)&monthly&\\
				&real\_c&real-estate climate index&monthly&\\
				&rpi&retail price index (year-on-year basis)&monthly&\\  
				&rrr&reserve requirement ratio&monthly&\\
				&shibor&Shanghai interbank offered rate (weighted averge)&monthly&\\
				&tax\_r&tax revenue&monthly&\\
				\hline
				\multirow{4}{*}{Global economy}
				&gold&global gold price (US dollarwise)&daily&World Gold Council\\
				&oil&crude oil price (West Texas Intermediate)&daily&St. Louis Fed\\
				&vix&S\&P500 volatility index&daily&WIND\\
				\hline
			\end{tabular}\label{tab:variable_summary}
		\end{center}
	\end{table}

	\begin{table}[htpb]
		\caption{Statistic descriptions}
		\setlength{\tabcolsep}{15pt}
		\small
		\begin{center}
			\begin{tabular}{lrrrrr}
				\hline
				Variable& 
				Mean& 
				Standard Dev. &
				Skewness&
				J.B. Statistic& Count\\
				\hline
				&&&&&\\
	       	ssec&1793.40&1039.33&1.08&1632.6&5686\\
			ssec\_r&0.09&2.65&12.15&48754295&5685\\
			sez&5793.46 &  4209.13&0.96&829.22&5686\\
			sez\_r&0.07  &   2.21 &0.96&45313&5685\\
			frx&7.39   &   0.85& -0.12&706.4&5686\\
			frx\_r&0.00  &    0.11 & 0.43&529747&5685\\
			dis\_r&3.80  &    1.84 & 2.40&466.99&257\\
			int\_r& 0.78   &   0.59 & 2.09&427.95&257\\
			bal\_g&4.83  &   3.80 & 0.25&0.94875&26\\
			cci&109.42   &   6.22 &  0.50&10.857&257\\
			cpi& 2.26  &    2.53 & 0.81&32.581&257\\
			comd\_g&42.75  &  10.56 & 0.67&2.803&26\\
			deb\_g&27.19   & 12.32 & 0.33&1.2994&26\\
			dom\_c&599448 &606773 & 1.18&60.85&257\\
			dom\_g&148.22  &  39.80 & 0.53&1.9997&26\\
			fix\_a&19.59   &   8.87 & 0.19&2.1097&257\\
		    frx\_res&16761 & 13999 & 0.19&30.582&257\\
			gdp\_p&28953 & 22391 &  0.57&2.7291&26\\
			gdp\_p2&8.22 &  1.99 & 0.82&3.4763&26\\
			gdp\_c&56.57  &  11.92 & 0.65&2.2989&26\\
			im\_ex&13.66 &    15.97 & 0.02&2.6891&257\\
			ind\_g&44.70  &   2.52 &-0.77&3.6808&26\\
			ind\_p&97.73    &  1.29 &  0.32&210&257\\
			ind\_v&11.46   &  4.27 & 0.21&10.587&257\\
			inf\_r&2.82  &   3.60  &2.22&75.228&26\\
		    mac\_c&96.44   &  3.31& -0.03&8.5258&257\\
			m2\_g&162.25  &  32.64 &-0.18&1.2309&26\\
			ppi&0.95   &   4.09 & 0.02&7.3328&257\\
			real\_c&101.28   &   3.63 &-0.58&15.889&257\\
			rpi&1.23    &  2.61 & 0.53&11.965&257\\
			rrr&12.88   &   5.20  &0.01&22.948&257\\
			shibor&3.39  &    2.73 & 2.32&434.42&257\\
			tax\_r&5061.64 &  4341.54 & 0.82&30.973&257\\
			gold&829.69  &  481.28&0.24&549.98&5686\\
			oil&54.46   &  28.87 & 0.46&329.34&5686\\
			vix& 19.90   &  8.14 & 2.08& 1766&5686\\
			\hline
			\end{tabular}\label{tab:data_stats}
		\end{center}
	\end{table}	
	
	\subsection{Data summary}
   Two prerequisites for sourcing data are (1) substantial variables are reserved in the a credible database, and (2) the variables have been studied and proven closely linked to predicting financial crises. Referring to \cite{KLR} study who have listed the most comprehensive table for crisis leading factors from various economic sectors, we similarly source $36$ variables, including $8$ daily, $17$ monthly and $11$ annually distributed factors corresponding to the currency and stock turbulence variations. For instance, in the national economy sector, external factors such as reserves and the real exchange rate, and financial factors such as domestic credit to GDP and interest rate, have already been proven capable of signalling the economic vulnerability in currency crashes for China \citep{PENG2008138}. Most variables are obtainable from WIND\footnote{WIND is a popular qualified financial database in mainland China. It contains both micro- and macro- economic variable data for researchers and practitioners. The only shortage is that the access is limited to commercial usage.} database, except discount rate and crude oil price from St. Louis Fed and gold price from World Gold Council (WGC). Time horizon covers from December $1^{st}$ of 1995 to December $1^{st}$ 2019, the most recent 24 years, which period is believed to contain sufficient crisis samples. Full data set is then split into 75\% in-sample and 25\% out-of-sample for both monthly and daily distributed data frames. 
   
   Table \ref{tab:data_stats} describes the key statistics for sourced variables on the full time length. From the columns of skewness and J.B. statistic, most of daily variables, such as stock index, exchange rates, interest rates and globe economy indicators, are fairly not normally distributed with sloppy values, which laterally attributes the market endogenous non-normality and exactly contributes to validate the EWS model predicting sensitivity on alienated outliers in financial series.

	\subsection{Identified crises}
   In this section, we first calculate the misspecification rate for each crisis classifier to compare the identifying robustness on different size samples, and then, similarly to previous studies, put the critical crisis events in chronological order to contrast the identified periods. The classifiers all listed in \ref{tab:crisis_classifier_formulae} will be implemented. In practice, the value for scaling coefficients of $\lambda_{fpi}$ and $\lambda_{cmx}$ are experimented\footnote{Gradient searching is implemented in the set of $\{1,1.5,2,2.5,3\}$ and the best performed value is taken with the minimal misspecification rate.} as $1.5$ and $2$ in the TQI classifier. The SWARCH classfier will be applied to stock with $2$ regimes and currency with $3$ regimes\footnote{The count of regimes $K$ is determined by calculating the RCM.}, respectively.
	
	\begin{table}[htpb]
		\caption{Compare classifiers' robustness: count of misspecified crisis observations.}
		\setlength{\tabcolsep}{5pt}
		\small
		\begin{center}
			\begin{tabular}{llrlrlr}
				\hline
				panel (a) : currency&\multicolumn{2}{c}{FPI}&\multicolumn{2}{c}{SWARCH}&\multicolumn{2}{c}{SWARCH$_{opt}$}\\
				&days&months&days&months&days&months\\
				\cline{2-7}
				Count of crises on full samples &25.0&48.0&1529&83&1517&83\\
				Count of crises on truncated full set&22.0&24.0&1092&52&1086&52\\
				Count of crises on test set &21.0&12.0&1034&55&1075&51\\
				Count of misspecified obs.&20.0&12.0&75.0&13.0&15.0&9.0\\
				$\%$ of misspecified obs.&1.05&13.9&3.96&15.1&0.791&10.5\\
				\hline
				Avg. count of misspecification&25.9&12.0&87.7&12.3&34.4&11.3\\
				Avg. $\%$ of misspecification &1.37&13.8&4.63&14.1&1.81&13.0\\
				\hline
				panel (b) : stock&\multicolumn{2}{c}{CMAX}&\multicolumn{2}{c}{SWARCH}&\multicolumn{2}{c}{SWARCH$_{opt}$}\\
				&days&months&days&months&days&months\\
				\cline{2-7}
				Count of crises on full samples &509&38.0&2097&148&2135&137\\
				Count of crises on truncated full set&172&15.0&641&25.0&653&21.0\\
				Count of crises on test set &202&21.0&646&37.0&640&27.0\\
				Count of misspecified obs.&30.0&6.00&25.0&12.0&25.0&6.00\\
				$\%$ of misspecified obs.&1.60&6.89&1.32&13.8&1.32&6.89\\
				\hline
				Avg. count of misspecification&53.5&10.1&85.5&9.07&39.7&8.32\\
				Avg. $\%$ of misspecification &2.86&11.6&4.51&10.4&2.09&9.56\\
				\hline
			\end{tabular}\label{tab:classifier_comp}
		\end{center}
	\end{table}

	\begin{table}[htpb]
		\caption{Identified crisis episodes for currency market.}
		\setlength{\tabcolsep}{7pt}
		\small
		\centering
		\begin{tabular}{p{120pt}p{100pt}p{80pt}p{80pt}}
			\hline
			critical event& evident sign&FPI&SWARCH$_{opt}$\\
			\hline
			&&&\\
			1997 Asian Crisis&Since the spring of 1996, Chinese yuan against U.S. dollar began to vibrate around 8.2 after deepening the 94's published `managed floating exchange rate market system' &$--$&1996.06-08; 1997.02-03; 1997.05,10\\
			\hline
			&&&\\
			2005 Exchange rate reform&On July 21, 2005, Chinese yuan appreciated 2.1\% overnight as China Central Bank announced to reform the exchange rate&2003.09-2004.09&2005.07-11 \\
			\hline
			&&&\\
			2007-08 Globe Financial Crisis&Starting from 2006, U.S. dollar to yuan continually depreciated 15\% till the mid-2008 China re-pegging yuan to U.S. dollar at 6.83 &2006.12-2007.12&2006.02,06; 2007.09-2009.06\\
			\hline
			&&&\\
			2010-11 Sovereign Debt Crisis&Since June of 2010, the exchange rate to U.S. dollar depreciated $2.49\%$ in a half year, till April of 2012 as China kept expanding the currency volatility range to $\pm 1.0\%$ &$--$&2010.06-12; 2012.05-10\\
			\hline
			&&&\\			
			2014 China managing float& The RMB currency volatility range was continually expanded to $\pm 2.0\%$ since 2014 March.
            &2014.06-12; 2015.03-07 &2014.03,04,09,12; 2015.03,04\\
			\hline
			&&&\\
			2015 `One basket currencies' scheme and Chinese stock crash&Starting from 2015 summer, the `stock-exchange resonance' led yuan against U.S. dollar dropped $5.8\%$over a year&$--$& 2015.08-12; 2016.03-12\\
			\hline
			&&&\\
			2018 Sino-US trading war&In 2018, U.S. dollar to yuan depreciated from 6.9 to 6.3 and then pulled back to 7 in 2019 spring&2017.05-12; 2018.03-06&2017.03-12; 2018.04-12; 2019.03-12\\
			\hline
		\end{tabular}\label{tab:identified_frx_crisis}
	\end{table}

	\begin{table}[htpb]
		\caption{Identified crisis episodes for stock market.}
		\setlength{\tabcolsep}{7pt}
		\small
		\centering
	    \begin{tabular}{p{120pt}p{100pt}p{80pt}p{80pt}}
			\hline
			critical event& evident sign&CMAX&SWARCH$_{opt}$\\
			\hline
			&&&\\
			1997 Asian Crisis&Starting from Jan. 1996, SSEC index swung up from 550 points to 1300 points over a year, then drastically dropped to 860 points at the start of 1997 and finally surged to 2500 points till 2000 Spring &$--$&1995.12-1998.05; 1998.08-1999.03; 1999.05-2000.05\\
			\hline
			&&&\\
			2001 State-owned Shares Reduction&SSEC index reduced 39.93\% in a half year after the `State-owned Shares Reduction' measure issued in June 2001&$--$&2001.07-10; 2002.05-10\\
			\hline
			&&&\\
			2004-05 Stock Market Reform& The average fall of A share prices for both SZSE and SSEC exceeded 11\% over the year of 2004, SSEC even drowned below 1000 points till June of 2005.&$--$&2004.04-07; 2005.02-06\\
			\hline
			&&&\\			
			2007-08 Globe Financial Crisis and 2010-11 Sovereign Debt Crisis&In 2008, the greatest drop of SSEC approximated to $70\%$, and in following 2010-2011, the bear market, the maximal decline was approaching $35\%$. & 2007.02-2009.02&2006.04-2011.02,09; 2012.02-04\\
			\hline
			&&&\\
			2013 `Cash Crunch'&The overall decline of SSEC was nearly $16\%$ in 2013, June &$--$&2013.06-09\\
			\hline
			&&&\\
			\multirow{1}{*}{2015 Chinese stock crash}&Since 2015 mid-June, SSEC twice plunged from climax 5187 to nearly below 3000 in two months&2014.12-2016.06&2014.11-2016.07\\
			\hline
			&&&\\
			\multirow{1}{*}{2018 Sino-US trading war}&Since mid-June 2018, SSEC depreciated more than 30\% in a few months &$--$&2018.06,10\\
			\hline
		\end{tabular}\label{tab:identified_stk_crisis}
	\end{table}
	
	The crisis classifier robustness will be measured in evaluating the stability of classified results as vary the crisis turmoil intensities for different pieces of observations. We implement all classifiers on full samples and test set to clarify their each performance by following the described procedure in the Section \ref{sec:misspecification_proc}. The test set for both non-boostrapping and boostrapping experiments is taken as one thirds of full samples. Table \ref{tab:classifier_comp} shows that, among the classifiers, the (average) misspecification rates given by the SWARCH$_{opt}$ classifier are smallest, $1.81\%$ and $13\%$ for currency and $2.09\%$ and $9.56\%$ for stocks, on daily and monthly identifications respectively. The classifier based on the SWARCH with the arbitrary cutoff, however, gives the most miserable misspecifation rates of $4.63\%$ and $14.1\%$ for currency on daily and monthly basis, and $4.51\%$ for stock daily basis. It gives $10.4\%$ for identifying monthly stocks, which is not worse than the CMAX brought value of $11.6\%$ but not enough to reverse its feebleness on crisis identification. In short, the crisis classifier built on the SWARCH with two-peak dynamically threshold is statistically proven to be the most robust one. In further comparisons, the SWARCH with arbitrary cutoff of 0.5 classifier will be ruled out to condense the distinction between the classic quantization index and the SWARCH methodologies.
	
	The `observable' financial crisis periods and their relevant starting sign evidences during the most recent 24 years from 1996 to 2019 will be chronologically displayed and visualized, as Table \ref{tab:identified_frx_crisis}, \ref{tab:identified_stk_crisis} and Figure \ref{fig:TQI_vs_SWARCH_mth} in Appendix II show, to investigate the credibility of crisis classifier's identifications in an intuitive way. 
	
    The highlighting session, such as 1997 Asian financial crisis, 2007-08 Global financial crisis, 2010-11 Sovereign Debt crisis and 2018 Sino-US trading war are cross-national financial crises for both markets, otherwise are national stylized critical facts, such as 2013 `Cash Crunch', 2015 Chinese stock market crash and macroeconomic control policy for specific market. In end of 90's Asian crisis, the currency market experienced seemly vibrations in the `managed floating' scheme, unlike the violent up and down stock index which heavily oscillated till 2000. In 2001, to deepen the enterprise reform, government decided to reduce the state-owned shares from stocks, which directly led an almost 40\% decline of SSEC in a half year. During 2004-2005, both markets were announced to be reformed, which made 2.1\% overnight depreciation for RMB against U.S. dollar and 11\% annual fall for A-shares price in both SSEC and SZSE. In the periods of notorious 2008-09 financial crisis and its follow-up 2010-11 Sovereign debt crisis, Chinese yuan devalued 15\% before pegging to the U.S. dollar and the bear market for stocks lasted for more than three years till the second half of 2012. In late May of 2013, the panic induced by inter-bank market capital shortage, alternatively called `Cash Crunch', spread to the stock market, which led SSEC a 16\% reduce in June. In 2014 Feb.,China took actions to guide the Chinese yuan weaker and swiftly turned to expand the volatility range into $\pm 2\%$ in March, which brought in Chinese yuan against U.S dollar drastic vibration in a few months. 2015 is the nightmare for Chinese stock investors. On June of 2015, after a chain of reactions between off-site allocation clean-up, on-site financing and deleveraging of graded funds, Chinese stock market was led to a severe crash that most stocks declined more than 50\% in twin-weeks, which touched off the fusing mechanism by imposing dual-thresholds of 5\% and 7\% on the index fusing benchmark. Even worse, a resonant effect was formed between the currency and stocks as  soon as the `one basket currencies' policy was published, which devaluate yuan against U.S. dollar by $5.8\%$. Starting from May 2017, China introduced ``counter-cyclical factor’’ \citep{Goldman2019} to reinforce market communication to the currency, and henceforth the exchange rate turned on ``roller coaster’’ mode, especially during the 2018 Sino-US trading war, yuan was devaluated by $11\%$ and hardly rallied to a strong level. The trade war also became the catalyst for stock market shocking decline of more than $30\%$ in a few months.

    Comparing identified results in Table \ref{tab:identified_frx_crisis} and \ref{tab:identified_stk_crisis}, the FPI and CMAX defined crisis classifiers show less sensitivity, even numbness, than SWARCH modeled volatility state jump classifier as match to the critical events. For instance, both FPI and CMAX sensors missed out 1997 Asian Crisis, SWARCH however sharply caught it and provided varied turmoil episodes given market-characteristic volatility dispersion. Without exception, SWARCH defined crisis classifier almost conquers all relative price dynamics variation between crisis and non-crisis periods. As for the detection precision, SWARCH classifier that includes the pre-crisis effect is superior to FPI and CMAX since all SWARCH model identified starting point is either earlier or more accurate than TQI index. For instance, the starting time for 2014 China managing floating is March, which time is precisely pinned by SWARCH but three months delayed by FPI, and the TQI inaccurate detection is embodied in identifying the 2005 Exchange rate reform, where FPI generates a dislocated period that yet covers the start point of the true turbulence. The visualized comparison as Figure \ref{fig:TQI_vs_SWARCH_mth} suggested makes the conclusion that SWARCH is more superior to traditional TQI index to sense the market turbulence more persuasive, as the volatile log returns can be perfectly reflected by SWARCH inferred filtering probabilities but scantly by FPI and CMAX. We hence reason that the SWARCH model with automatically optimized thresholds is not only statistically convincing among crisis classifiers in crisis identification for both daily and monthly predictions, but evidently persuasive as match to the historic critical events that lead severe market turmoils in reality.

	\subsection{Forecasting power}
	The predicted results will be displayed in two aspects of the hit-ratio calculations, including the correct predictions, correctly called onsets, false alarms and predicted days in advance, and the statistical calibration metrics of QPS, Youden $J$ and SAR.  Table \ref{tab:model_power_currency} and \ref{tab:model_power_stock} give the calculated values on out-of-samples in terms of daily (short-term) and monthly (long-term) data frames for currency and stock markets respectively, where the best performed metrics are bold-type in each row to present the corresponding combined EWS model performance.
	
	Comparing the results in sub-panels of (a)/(b)-1 and (a)/(b)-2, we verify that SWARCH classifier can truly boost the EWS performance in terms of either higher predicting precision or lower false calling alarms especially as combine econometric models of logistic regressions. For instance, the FPI combined LR gives $61.5\%$ correctly called crisis observations and $28.1\%$ false alarm rate while the SWARCH hybrid LR improves the values by $20.8\%$ (to be $82.3\%$ correctly predicted crises) and $28.1\%$ (to be $0.00\%$ false alarms) for currency market short-term prediction. SWARCH shares a similar boosting on KLR, except for stock long-term prediction, neither the hit ratios for predicted crisis observations and predicted crisis onsets nor the statistical metrics for Youden $J$ and SAR get significant improvements. Among the machine learning techniques, the classification tree models of RF and XGBoost can benefit from SWARCH classifier by diminishing the false alarms. For example, for the short-term prediction on currency market, the percentage of false alarms produced by FPI combined RF model is $23.1\%$ while being decreased to be $0.00\%$ as substituting the SWARCH classifier. Besides, the SWARCH combined EWS model will provide longer early warning periods especially for long-term predictions. The typical evidence can be found as compare the row of hit(4) between (b)-1 and (b)-2 in Table \ref{tab:model_power_currency}, that the called advanced days for crisis onsets have been prolonged for 1-2 months, which implies ample reaction time will be suggested to regulators or decision makers to respond the market vulnerability.

	\begin{table}[htpb]
		\caption{Compare EWS models predictive power: hit-ratios and score metrics for currency.}
		\setlength{\tabcolsep}{8pt}
		\small
		\begin{center}
			\begin{tabular}{lrrrrrr}
				\hline
				\multirow{2}{*}{panel (a): short-term}&&&&&&\\
				&LR&KLR&NN&RF&XGBoost&AttnLSTM\\
				(a)-1: FPI &&&&&&\\
				\cline{1-1}
				hit(1)$^{a}$&61.5&74.3&94.8&\textbf{100}&94.8&97.4\\
				hit(2)$^{b}$&51.0&71.8&84.4&59.3&22.0&\textbf{86.8}\\
				hit(3)$^{c}$&28.1&38.9&29.4&23.1&\textbf{0.00}&29.2\\
				hit(4)$^{d}$&2.57&2.51&2.92&2.81&\textbf{3.13}&2.45\\
				QPS&0.209&0.501&0.00$^{*}$&0.00$^{*}$&0.00$^{**}$&0.00$^{*}$\\
				Youden $J$&0.498&0.432&0.773&\textbf{0.993}&0.948&0.899\\
				SAR&0.692&0.591&0.789&\textbf{0.834}&0.679&0.798\\
				\hline
				(a)-2: SWARCH&&&&&&\\
				\cline{1-1}
				hit(1)&82.3&81.7&96.8&\textbf{100}&\textbf{100}&\textbf{100}\\
				hit(2)&45.5&55.3&84.5&0.00&54.5&\textbf{90.9}\\
				hit(3)&0.00&32.9&\textbf{0.00}&\textbf{0.00}&\textbf{0.00}&\textbf{0.00}$^{**}$\\
				hit(4)&3.07&2.39&\textbf{3.18}&0.00&3.13&2.32\\
				QPS&-0.17&-0.254&0.00$^{*}$&0.00$^{**}$&0.00$^{**}$&0.00$^{*}$\\
				Youden $J$&0.823&0.159&0.986&\textbf{1.00}&\textbf{1.00}&0.969\\
				SAR&0.865&0.646&0.792&\textbf{1.00}&0.667&\textbf{0.896}\\
				\hline
				\multirow{2}{*}{panel (b): long-term}&&&&&&\\
				&LR&KLR&NN&RF&XGBoost&AttnLSTM\\
				(b)-1: FPI &&&&&&\\
				\cline{1-1}
				hit(1)&66.7&70.8&91.7&\textbf{100}&83.3&95.8\\
				hit(2)&0.00&\textbf{100}&50.0&0.00&0.00&50.0\\
				hit(3)&15.0&36.8&14.8&12.5&\textbf{4.17}&7.41\\
				hit(4)&0.00&\textbf{2.20}&1.00&0.00&0.00&1.00\\
				QPS&-0.129&-0.167&0.097&0.129&\textbf{-0.096}&0.105\\
				Youden $J$&0.561&0.290&0.785&\textbf{0.895}&0.807&0.837\\
				SAR&0.723&0.612&0.736&0.758&0.734&0.782\\
				\hline
				(b)-2: SWARCH&&&&&&\\
				\cline{1-1}
				hit(1)&84.6&95.8&\textbf{100}&\textbf{100}&\textbf{100}&98.0\\
				hit(2)&88.9&\textbf{100}&88.9&88.9&88.9&87.5\\
				hit(3)&18.5&33.8&\textbf{0.00}&\textbf{0.00}&\textbf{0.00}&\textbf{0.00}\\
				hit(4)&2.46&2.5&2.8&\textbf{2.9}&\textbf{2.9}&2.4\\
				QPS&0.062&0.254&0.00$^{*}$&0.00$^{**}$&0.00$^{**}$&0.00$^{*}$\\
				Youden $J$&0.581&0.127&0.923&\textbf{1.00}&\textbf{1.00}&0.934\\
				SAR&0.769&0.679&0.889&0.889&0.667&\textbf{0.897}\\
				\hline
			\end{tabular}\label{tab:model_power_currency}
		\end{center}
		
		\footnotesize{$^{a}$ $\%$ of correctly called crisis observations.}\\
		\footnotesize{$^{b}$ $\%$ of correctly called onsets.}\\
		\footnotesize{$^{c}$ $\%$ of false alarms.}\\
		\footnotesize{$^{d}$ count of alarming advanced days.}\\
		\footnotesize{$^{*}$,$^{**}$ refer to with $5\%$ and $1\%$ significance level respectively.}
	\end{table}
	
	In addition, the EWS model based on machine learning predictive techniques outperforms that based on traditional econometric models with better goodness-of-fit and statistical scores especially as the SWARCH classifier joins. The average hit ratio of correct predicted crises extracted from NN, RF, XGBoost and AttnLSTM based EWS is greater than $90\%$ for both markets predictions (except the short-term prediction for stock with CMAX classifier), which wins the LR and KLR that model hardly get more than $90\%$ correct predictions with a commanding lead. Furthermore, most zeros of QPS and ones of Youden $J$ metric values are given by SWARCH combined machine learning models, which also strongly validates the predicting robustness of these hybrid state-of-art brain-like learning models. In fact, it is quite differential to specify machine learning models forecasting power, for instance, even though classification tree models of RF and XGBoost bring in $100\%$ predicting precision on crisis observations and $0.00\%$ false alarms (as refer to the hit (1) and hit (3) in (a)-2 and (b)-2 sub-panels for both markets), the neural networks of NN and AttnLSTM generally perform better on predicting crisis onsets and advanced days (as refer to hit (2) and hit (4) in (a)-2 and (b)-1 of Table \ref{tab:model_power_currency} and (a)/(b)-2 of Table \ref{tab:model_power_stock}).
	
    In contrast to previous study on comparing econometric models constructed and machine learning based warning system for predicting banking crisis that claims conventional models already fairly efficient to use available information \citep{BEUTEL2019}, we conclude a different pattern in line with the measured results that the machine learning based predictive models, especially the deep learning based EWS, will not lose either credibility or robustness on out-of-sample predictions given sufficient endogenous and exogenous features data and SWARCH model inferences. 
	
	\begin{table}[htpb]
		\caption{Compare EWS models predictive power: hit-ratios and score metrics for stock.}
		\setlength{\tabcolsep}{8pt}
		\small
		\begin{center}
			\begin{tabular}{lrrrrrr}
				\hline
				\multirow{2}{*}{panel (a): short-term}&&&&&&\\
				&LR&KLR&NN&RF&XGBoost&AttnLSTM\\
				(a)-1: CMAX &&&&&&\\
				\cline{1-1}
				hit(1)&49.1&47.2&79.2&63.2&56.6&\textbf{86.8}\\
				hit(2)&\textbf{50.0}&\textbf{50.0}&\textbf{50.0}&33.3&0.00&\textbf{50.0}\\
				hit(3)&43.1&46.4&36.1&\textbf{0.00}&\textbf{0.00}&44.4\\
				hit(4)&\textbf{2.71}&2.43&2.00&1.80&0.00&2.5\\
				QPS&0.00$^{*}$&-0.10&0.00$^{*}$&-0.07&-0.07&0.00$^{*}$\\
				Youden $J$&0.038&0.028&0.770&0.132&0.056&\textbf{0.836}\\
				SAR&0.643&0.656&0.581&0.102&\textbf{0.718}&0.663\\
				\hline
				(a)-2: SWARCH&&&&&&\\
				\cline{1-1}
				hit(1)&92.1&53.2&96.3&\textbf{100}&\textbf{100}&\textbf{100}\\
				hit(2)&63.2&61.9&47.4&47.4&47.4&\textbf{89.5}\\
				hit(3)&38.7&50.4&\textbf{0.00}&\textbf{0.00}&\textbf{0.00}&0.00$^{*}$\\
				hit(4)&2.74&2.47&\textbf{3.21}&3.15&3.15&2.42\\
				QPS&1.14&0.08&0.00$^{*}$&\textbf{0.00}$^{**}$&\textbf{0.00}$^{**}$&0.00$^{*}$\\
				Youden $J$&0.218&0.182&0.963&\textbf{1.00}&\textbf{1.00}&0.965\\
				SAR&0.554&0.548&0.821&0.712&0.767&\textbf{0.855}\\
				\hline
				\multirow{2}{*}{panel (b): long-term}&&&&&&\\
				&LR&KLR&NN&RF&XGBoost&AttnLSTM\\
				(b)-1: CMAX &&&&&&\\
				\cline{1-1}
				hit(1)&59.3&76.3&\textbf{100}&93.3&60.0&\textbf{100}\\
				hit(2)&0.00&50.0&\textbf{100}&0.00&0.00&\textbf{100}\\
				hit(3)&40.6&48.2&11.1&12.0&\textbf{0.00}&11.1\\
				hit(4)&0.00&\textbf{1.50}&1.00&0.00&0.00&\textbf{1.50}\\
				QPS&0.813&0.180&\textbf{0.102}&0.610&-0.203&0.111\\
				Youden $J$&0.454&0.605&\textbf{0.932}&0.502&0.60&0.923\\
				SAR&0.549&0.712&0.805&0.589&0.739&\textbf{0.834}\\
				\hline
				(b)-2: SWARCH&&&&&&\\
				\cline{1-1}
				hit(1)&84.2&73.0&\textbf{100}&\textbf{100}&\textbf{100}&\textbf{100}\\
				hit(2)&0.00&50.0&\textbf{100}&0.00&0.00&\textbf{100}\\
				hit(3)&39.3&26.2&20.0&\textbf{0.00}&\textbf{0.00}&33.3\\
				hit(4)&0.00&2.40&1.5&0.00&0.00&\textbf{2.50}\\
				QPS&0.277&-0.127&0.185&\textbf{0.00}$^{**}$&\textbf{0.00}$^{**}$&\textbf{0.00}$^{*}$\\
				Youden $J$&0.581&0.442&0.870&\textbf{1.00}&\textbf{1.00}&0.942\\
				SAR&0.678&0.713&0.852&\textbf{1.00}&0.667&0.827\\
				\hline
			\end{tabular}\label{tab:model_power_stock}
		\end{center}
	\end{table}
	
	\begin{sidewaystable}[htpb]
		\caption{Model selection for currency and stock markets.}
		\setlength{\tabcolsep}{3pt}
		\small
		\begin{center}
			\begin{tabular}{lrrrr}
				\hline
				market &\multicolumn{2}{c}{currency}&\multicolumn{2}{c}{stocks}\\
				classifier&\multicolumn{1}{c}{FPI}&\multicolumn{1}{c}{SWARCH}&\multicolumn{1}{c}{CMAX}&\multicolumn{1}{c}{SWARCH}\\
				\hline
				&&&&\\
				panel (a): short-term &&&&\\
				hit(1)$>90$&NN,RF,XGBoost,AttnLSTM&NN,XGBoost,AttnLSTM&--&LR,NN,RF,XGBoost,AttnLSTM\\
				hit(2)$>80$&NN,AttnLSTM&NN,AttnLSTM&--&AttnLSTM\\
				hit(3)$<10$&--&LR,NN,RF,XGBoost,AttnLSTM&RF,XGBoost&NN,RF,XGBoost,AttnLSTM\\
				hit(4)$>3$ days&XGBoost&LR,NN,XGBoost&--&NN,RF,XGBoost\\
				$|\text{QPS}|<0.05$&NN,RF,XGBoost,AttnLSTM&NN,RF,XGBoost,AttnLSTM&LR,NN,AttnLSTM&NN,RF,XGBoost,AttnLSTM\\
				Youden $J>0.9$&RF,XGBoost&NN,RF,XGBoost,AttnLSTM&--&NN,RF,XGBoost,AttnLSTM\\
				SAR$>0.8$&RF&LR,RF,AttnLSTM&--&AttnLSTM\\
				\hline
				best model&\multicolumn{2}{c}{SWARCH-NN/AttnLSTM}&\multicolumn{2}{c}{SWARCH-AttnLSTM}\\
				\hline
				&&&&\\
				panel (b): long-term &&&&\\
				hit(1)$>90$&NN,RF,AttnLSTM&KLR,NN,RF,XGBoost,AttnLSTM&NN,RF,AttnLSTM&NN,RF,XGBoost,AttnLSTM\\
				hit(2)$>80$&KLR&LR,KLR,NN,RF,XGBoost,AttnLSTM&NN,AttnLSTM&NN,AttnLSTM\\
				hit(3)$<10$&XGBoost,AttnLSTM&NN,RF,XGBoost,AttnLSTM&XGBoost&RF,XGBoost\\
				hit(4)$>2$ month&KLR&LR,KLR,NN,RF,XGBoost,AttnLSTM&--&KLR,AttnLSTM\\
				$|\text{QPS}|<0.05$&--&NN,RF,XGBoost,AttnLSTM&--&RF,XGBoost,AttnLSTM\\
				Youden $J>0.9$&--&NN,RF,XGBoost,AttnLSTM&NN,AttnLSTM&RF,XGBoost,AttnLSTM\\
				SAR$>0.8$&--&NN,RF,AttnLSTM&NN,AttnLSTM&NN,RF,AttnLSTM\\
				\hline
				best model&\multicolumn{2}{c}{SWARCH-NN/RF/AttnLSTM}&\multicolumn{2}{c}{SWARCH-AttnLSTM}\\
				\hline
			\end{tabular}\label{tab:opt_model_selection}
		\end{center}
	\end{sidewaystable}
	
    To concentrate the model comparisons in a compact way, we further clarify a series of refining conditions\footnote{The refining conditions are statistical metrics with appropriate threshold level to rule out unsatisfied models. The threshold value can be country or market personalized in practice.} into a unified system. As Table \ref{tab:opt_model_selection} shows, the selection standards are listed at the left side and models that can pass through each selection condition will then be screened till the the final row, the most frequently screened model that is picked as the best EWS for predicting corresponding market and required term displays. According to final model options, SWARCH crisis classifier is indisputably preferred to both markets for both short- and long- term prediction. The market however shows idiosyncrasy in predictive model selection. Particularly, the stock market is solely partial to the AttnLSTM, while the currency market seems to prefer both of NN and AttnLSTM for short-term prediction and extra includes the RF for long-term prediction. To avoid ambiguous model determinations, further comparisons for currency market are required. Back to Table \ref{tab:model_power_currency}, AttnLSTM and RF produced more bold-labelled results for short-term panel (refer to (a)-2) and long-term panel (refer to (b)-2) respectively. Thus in brief, the unified comparison suggests the SWARCH-AttnLSTM as the best option for predicting currency and stock crises, particularly yet for long-term currency crises prediction, recommends the SWARCH-RF.
    
	\subsection{Back testing and Reality check}
	The predictive models in the financial applications that are blamed for its lack of practical verification bear the questioning on whether the predictability is equivalent to the effectiveness in real world. Thus the dynamical portfolio allocation will be applied according to investors' risk aversion variation given properly predicted the crisis episodes, thereby to validate whether the prominent crisis forecasting system will bring attractive returns in practice. 
	\begin{table}[htpb]
		\caption{Back-test results of asset allocation on out-of-samples.}	
		\setlength{\tabcolsep}{2pt}
		\centering
		\small
		\begin{tabular}{llrrrrrr}
			\hline
			risk aversion level&&\multicolumn{2}{c}{low: $\gamma = 1$}&
			\multicolumn{2}{c}{medium: $\gamma = 2$} & 
			\multicolumn{2}{c}{high: $\gamma = 3$}
			\\ \hline
			&&&&&&&\\
			panel(a): daily&&Sharpe ratio&CER&Sharpe ratio&CER&Sharpe ratio&CER\\ 
			\hline
			\multirow{3}{*}{benchmark}&buy-and-hold&0.236&7.74&0.241&-2.66&0.244&-6.12\\
			&TQI&0.307&8.42&0.306&0.204&0.291&-3.67\\
			&SWARCH&0.454&10.2&0.442&2.25&0.390&-2.82\\
			\hline
			\multirow{6}{*}{TQI - }&-LR&0.344& 8.04&0.302&1.33&0.288&-2.36\\
			&-KLR&0.234&7.76&0.238&-0.943&0.243&-3.37\\
			&-NN&0.362&7.83&0.312&1.65&0.295&-2.24\\
			&-RF&0.351&7.84&0.305&1.40&0.289&-2.31\\
			&-XGBoost&0.336&7.85&0.340&0.893&0.344&-2.34\\
			&-AttnLSTM&0.368&8.67&0.372&1.43&0.391&-2.28\\
			\hline
			\multirow{6}{*}{SWARCH -}&-LR&0.498&10.8&0.441&2.92&0.391&0.015\\
			&-KLR&0.504&9.31&0.440&2.86&0.389&0.010\\
			&-NN&0.517&11.8&0.443&2.89&0.391&0.020\\
			&-RF&0.508&11.0&0.442&2.89&0.390&0.004\\
			&-XGBoost&0.508&10.5&0.442&2.89&0.391&0.008\\
			&-AttnLSTM&0.518&11.5&0.461&3.22&0.407&0.024\\
			\hline
			&&&&&&&\\
			panel(b): monthly&&Sharpe ratio&CER&Sharpe ratio&CER&Sharpe ratio&CER\\ 
			\hline
			\multirow{3}{*}{benchmark}&buy-and-hold&0.409&20.3&0.448&-14.1&0.478&-26.3\\
			&TQI&0.448&22.2&0.469&-15.6&0.481&-27.6\\
			&SWARCH&0.497&22.4&0.502&-10.8&0.509&-20.5\\
			\hline
			\multirow{6}{*}{TQI - }&-LR&0.517&24.5&0.495&-3.65&0.496&-14.3\\
			&-KLR&0.505&23.8&0.487&-2.27&0.503&-13.6\\
			&-NN&0.534&27.2&0.455&-2.92&0.462&-15.8\\
			&-RF&0.492&22.4&0.499&-0.07&0.508&-11.3\\
			&-XGBoost&0.493&22.5&0.499&-0.07&0.509&-11.1\\
			&-AttnLSTM&0.541&26.8&0.519&1.57&0.526&-10.3\\
			\hline
			\multirow{6}{*}{SWARCH -}&-LR&0.503&22.6&0.505&0.350&0.506&-11.4\\
			&-KLR&0.519&25.8&0.492&-0.741&0.497&-15.8\\
			&-NN&0.516&23.9&0.509&0.473&0.512&-11.0\\
			&-RF&0.513&23.8&0.512&0.782&0.513&-11.3\\
			&-XGBoost&0.513&23.8&0.512&0.781&0.514&-11.2\\
			&-AttnLSTM&0.556&29.8&0.534&2.24&0.528&-10.5\\
			\hline
		\end{tabular}
		\label{tab:backtest}
	\end{table}
	
	In the study, three assets of SSEC index, exchange rate against U.S. dollar and risk-free interest\footnote{The risk-free rate is adopted as the bank short-term deposit interest rate.} will be accounted into the back test. To make a fair situation, we impose a key assumption that no informative asymmetry exists as practitioners access the early warning model outputs and construct the portfolio based on the produced warning signals varied risk aversion. Three benchmark portfolios will be constructed by taking `buy-and-hold', solely applied `TQI' and solely applied `SWARCH' strategies that not involve predicted results. 
	
	Table \ref{tab:backtest} shows the Sharpe ratios and Certainty Equivalent Return (CER) under different constant base risk aversion levels of low, medium and high for both daily and monthly exercising scenarios. Among three benchmarks, the crisis classifier adopted portfolios, especially the SWARCH classifier adopted, is most preferred by either doubling (for short-term daily panel) or gaining a quarter (for long-term monthly panel) than the low risk aversion buy-and-holders produced Sharpe ratios. In panel (a), as compare to buy-and-hold, the TQI based EWS though contributes the investors by gaining Sharpe ratio, the gain is in a mild range of no more than $50\%$, in contrast, SWARCH based forecasting system boosts the gain to more than $100\%$. Further referring to the CER, SWARCH hybrid EWS even remarkably turns the value from negative to positive (even though close to zero) in the high risk aversion column. The situation for monthly case, as panel (b) shows, however, is quite different since neither the TQI based nor SWARCH based EWS models involved portfolios returns on Sharpe ratios and CERs are significantly improved as compare to benchmarks. But it is worthy to note, as the risk aversion is in medium level, the positive CER can only be retrieved by SWARCH based EWS and TQI-AttnLSTM models, which advocates the Attention based deep learning model robustness in gaining the practical benefits. In short, SWARCH-AttnLSTM is highly recommended to short-term investors regardless of the risk preference since it universally issues the greatest Sharpe ratios of $0.518$, $0.461$ and $0.407$, and maximal CERs, except SWARCH-NN gives $11.8$ given $\gamma=1$, of $3.22$ and $0.024$ among all calculations, and it still wins in the long-term investment but with a weak edge. 
	
	We also find the EWS may reverse investors earning profit situation according to their risk aversion level. Particularly, the high-risk seeker ($\gamma=1$) who previously have lower Sharpe ratios than conservative investors ($\gamma=3$) as holding benchmark portfolios will benefit more as apply the EWS to adjust portfolio weights especially for short-term investment. This conclusion is consistent with \cite{HASLER2019182} who claim that accounting for timing of equity risks will advantage the investing portfolio returns by increasing the CER of $3\%$. Meanwhile, since the selling pressure of short-term investors is proven positively linked to market crash \citep{Fan2020}, the dynamically constructed portfolio according to EWS produced warning signals will faithfully support practical solution to adjust asset weights and to further avert dreadful panic that always manifests in uninstructed situation.

	To investigate whether the EWS forecasting robustness in back-testing is accidental or not, reality check will be implemented. The benchmarks mainly keep same settings as back-testing, but will be slightly changed in opting the EWS contrast group, that is the benchmark of TQI crisis classifier will only appoint to TQI based EWS models group and vice versa. This setting will project the predictive models effort more in comparisons. 
	
	Table \ref{tab:model_rc} reasons the credibility of back-testing results given p-values below $0.1$, which indicates the null hypothesis that the EWS does not help to diminish the return variance can be strongly rejected. The p-values that exceed $0.1$ or close to $0.1$, mainly locate in TQI-XGBoost based EWS for daily prediction and traditional econometric models of LR and KLR based EWS for monthly prediction, which cautions investors against such EWS models that may produce less reliable results to minimize the volatility triggered risks in returns. Besides, the SWARCH combined machine learning EWS models, especially combined tree models of RF, are most likely to pass through the reality check by harvesting more zero or close to zero p-values in the comparison. The highly recommended model in back-testing, SWARCH-AttnLSTM, however puts a bit of damper by losing credibility at $0.05$ level, specifically providing $0.054$, $0.049$ and $0.062$, as compare to the benchmark of solely adopting SWARCH classifier in terms of daily prediction. In the comparisons, the suspect on the credibility of composing a hybrid warning system in contrast to solely exercising identified crises from classifiers, though, cannot be fully dislodged since most presented $p_{B1}$'s are generally smaller than $p_{B2}$'s, which implies the return variance harder to be kept lowering as predictive model is introduced after identified crisis being applied to ameliorate the asset returns, the machine learning hybrid EWS models adopted on monthly prediction strike the this suspicious assertion by giving smaller $p_{B2}$ than $p_{B1}$, for example SWARCH-NN/RF/XG present $p_{B2}$ zero at $0.1\%$ significant level but $p_{B1}$ greater than $0.01$ for low and medium risk aversion investors.
	
	In general, the SWARCH combined machine learning models generally outperform among the developed hybrid EWS models, and such stylized models show comparative practical persuasion in varied investing risk preference. Traditional econometric models, especially combined with the classic technically quantified index crisis classifier, seem to faintly satisfy modern requisite on not only the high predicting precision but the demanding portability in practice. Hence, on the premise that both data quantity and quality can be guaranteed, the investors or decision makers are more likely to appeal to the stylized warning systems instructed warning signals.
	
	\begin{table}[t] 
		\caption{Reality check on EWS model robustness.}
		\setlength{\tabcolsep}{8pt}
		\small
		\begin{center}
			\begin{tabular}{llcccccc}
				\hline
				risk avresion level&&\multicolumn{2}{c}{low: $\gamma=1$}&\multicolumn{2}{c}{medium: $\gamma=2$}&\multicolumn{2}{c}{high: $\gamma=3$}\\
				\hline
				&&&&&&&\\           
				panel (a): daily&&$p_{B_{1}^{a}}$&$p_{B_{2}^{a}}$&$p_{B_{1}}$&$p_{B_{2}}$&$p_{B_{1}}$&$p_{B_{2}}$\\
				\hline
				\multirow{6}{*}{TQI-}&-LR&0.039&0.054&0.016&0.084&0.021&0.083\\
				&-KLR&0.031&0.110&0.029&0.091&0.021&0.086\\
				&-NN&0.060&0.090&0.047&0.111&0.036&0.102\\
				&-RF&0.001&0.011&0.005&0.012&0.010&0.011\\
				&-XG&0.087&0.125&0.079&0.092&0.078&0.083\\
				&-AttnLSTM&0.051&0.081&0.043&0.082&0.037&0.072\\
				\hline
				\multirow{6}{*}{SWARCH-}&-LR&0$^{***}$&0.073&0.002&0.066&0.003&0.061\\
				&-KLR&0.013&0.087&0.032&0.072&0.034&0.088\\
				&-NN&0$^{***}$&0.083&0.002&0.078&0.002&0.072\\
				&-RF&0$^{***}$&0$^{***}$&0.002&0$^{***}$&0.002&0$^{***}$\\
				&-XG&0$^{***}$&0.026&0.002&0.017&0.002&0.023\\
				&-AttnLSTM&0$^{***}$&0.054&0.001&0.049&0.002&0.062\\
				\hline   
				&&&&&&&\\    		   
				panel (b): monthly&&$p_{B_{1}^{a}}$&$p_{B_{2}^{a}}$&$p_{B_{1}}$&$p_{B_{2}}$&$p_{B_{1}}$&$p_{B_{2}}$\\
				\hline
				\multirow{6}{*}{TQI-}&-LR&0.070&0.156&0.071&0.159&0.071&0.121\\
				&-KLR&0.050&0.171&0.064&0.124&0.049&0.133\\
				&-NN&0.031&0.066&0.032&0.067&0.036&0.067\\
				&-RF&0.018&0.014&0.016&0.015&0.008&0.013\\
				&-XG&0.017&0.014&0.019&0.020&0.008&0.010\\
				&-AttnLSTM&0.019&0$^{***}$&0.024&0$^{***}$&0.006&0$^{***}$\\
				\hline
				\multirow{6}{*}{SWARCH-}&-LR&0.019&0.091&0.017&0.083&0.007&0.075\\
				&-KLR&0.057&0.079&0.054&0.076&0.056&0.092\\
				&-NN&0.020&0$^{***}$&0.024&0$^{***}$&0.006&0$^{***}$\\
				&-RF&0.020&0$^{***}$&0.025&0$^{***}$&0.008&0$^{***}$\\
				&-XG&0.019&0$^{***}$&0.021&0$^{***}$&0.008&0.014\\
				&-AttnLSTM&0.014&0.009&0.023&0.002&0.010&0.007\\
				\hline   
			\end{tabular}\label{tab:model_rc}
		\end{center}
		
		\footnotesize{$^{a}$ $B_{1}$ is the benchmark of buy-and-hold. $B_{2}$ is TQI for TQI based EWS and SWARCH for SWARCH based EWS.}\\
		\footnotesize{$^{***}$ denote at $0.1\%$ significance level.}
	\end{table}
	
	\subsection{Leading factors}
    The estimation on leading factors will not only attribute the contributing degree to inform of predicting crises, also distinguish the factors that should be regularly inspected among multi-source information as well. For each applied predictive model, the input features importance can be estimated and extracted from model produced results, specifically, the logistic regression bases on the parameter's coefficient, non-parametric KLR approach relies on the calculation for noise-to-signal ratios, neural networks have dropping one factor each time to infer the accuracy loss as feature importance, random forest hires the metric of mean decrease accuracy (MDA)\footnote{There are in fact two metrics for RF to draw the feature importance, mean decrease accuracy (MDA) and mean decrease Gini (MDG). In the study, we take MDA since its observed value variation is more distinctive than MDG provided in terms of information gain.}, XGBoost uses the gain that represents the fractional contribution of each factor based on the total gain of this factor's split, and the attention based LSTM extracts the final learnt attention weights for each features.
    
    Table \ref{tab:leading_summary} further summarizes the model detected leading factors by collecting the estimated results on factor contribution that are displayed in Table \ref{tab:leading_currency_short}-\ref{tab:leading_stock_long}, where the highlighted bold numbers are recognized significant impact on the crisis prediction. According to the Table \ref{tab:leading_summary}, the most frequently drawn factors are frx (exchange rate), gdp\_p2 (per capital GDP year-on-year basis) and ind\_g (industrial added value as the percentage of annual GDP), which suggests that currency exchange rate dynamics itself and macroeconomic factors relating to the national production situation should be early cautioned for crisis prediction regardless of the prediction term. In fact, the real exchange rate has been proven to be the most significant leading factor for predicting currency crisis \citep{KLR,BERG1999,BABECKY20131}, and meanwhile provides the significant evidence to be contagious to the stock market \citep{CHATZIS2018353}, which factor also gets accreditation in our study. It is noted that, the reverse contagion relationship from stock to currency seems to not exist since the neither stock index dynamics nor stock return rate significantly contribute to currency crisis prediction. Different from previous argument \citep{KLR,BERG1999} that claims the reserves and exports are crucial to indicate currency crisis, we find rare connection between such factors (frx\_res and im\_ex) and Chinese currency and stock market crisis forewarning, in contrary, the domestic economy related indicators (gdp\_p2 and ind\_g) are highlighted. It implies that the market turbulence is more likely to be prepended by the forthcoming economy recession, not be fully signed by the monetary policy control and international trades. Furthermore, the leading factors that precede to the market crisis are essentially different, in some case, even market-characteristic, for instance, the inflation factors, such as discount rate and CPI, are hardly ignored in signing the long-term currency crisis but do not share any contribution to warn stocks. This may imply the domestic money purchasing power could drive the volatile level of foreign exchange market in the fiscal policy channel, but seems not possibly to rock the share prices. 

    \begin{table}[htpb]
         \centering
      	\caption{Leading factors: summary of model detected leading factors for currency and stock market.}
		\setlength{\tabcolsep}{7pt}
		\small
        \begin{tabular}{lllll}
            \hline
            &\multicolumn{2}{l}{currency}  & \multicolumn{2}{l}{stocks} \\
            \cline{2-5}
                 &short-term& long-term & short-term & long-term \\
                 \hline
                 &&&&\\
	            ssec&&&(1)RF&(4)NN,RF,XG,Attn\\
				ssec\_r&&&(2)XG,Attn&\\
				sez&&&(2)RF,XG&\\
				sez\_r&&&&\\
				frx&(3)KLR,NN,RF&(3)KLR,NN,Attn&(2)NN,Attn&(3)LR,KLR,Attn\\
				frx\_r&(4)NN,RF,XG,Annt&&&\\
				frx\_res&&&&\\
				dis\_r&&(3)NN,RF,XG&&\\
				int\_r&&&(1)LR&\\
				bal\_g&&(2)KLR,RF&(2)XG,Attn&(3)KLR,RF,Attn\\
				cci&(2)KLR, Attn&&&\\
				cpi&&(1)XG&&\\
				comd\_g&(1)LR&&(2)KLR,Attn&\\
				deb\_g&(1)LR&(1)LR&(2)LR,KLR&\\
				dom\_c&&(2)RF,XG&&\\
				dom\_g&(1)LR&&(3)KLR,NN,Attn&\\
				fix\_a&&&&\\
				gdp\_p&&(1)KLR&&(1)KLR\\
				gdp\_p2&(2)LR,NN&(2)RF,XG&(2)KLR,Attn&(1)RF\\
				gdp\_c&&(1)NN&&\\
				im\_ex&&&&\\
				ind\_g&(2)LR,NN&(2)NN,Attn&(3)LR,NN,Attn&(1)LR\\
				ind\_p&(3)KLR,XG,Attn&&&\\
				ind\_v&&&&\\
				inf\_r&(1)KLR&&(1)LR&(4)NN,RF,XG,Attn\\
				mac\_c&(3)LR,KLR,RF&&&\\
				m2\_g&&&(1)KLR&\\
				ppi&(2)RF,XG&&(3)NN,XG,Attn&\\
				real\_c&(1)LR&&(1)XG&\\
				rpi&&&(2)LR,XG&\\
				rrr&(1)KLR&&&\\
				shibor&&&&\\
				tax\_r&&&&\\
				vix&(2)XG,Attn&&(2)RF,Attn&\\
				gold&&&(3)NN,RF,XG&\\
				oil&&&&\\
            \hline
        \end{tabular} \label{tab:leading_summary}
    \end{table}
    
    The summarized table information also instructs short-term and long-term investments in inspecting the leading factors with varied perspectives. On one hand, the return rates more influence than the price/index itself on short-term prediction, for instance, XGBoost and Attention-LSTM catch the ssec\_r, and neural networks, random forest, XGBoost and Attention-LSTM catch the frx\_r to predict stocks and currency market in short-term, but disappear to hint the long-term crisis. On the other hand, the short-term prediction requires more government published composite factors from external economy sectors, such as the macroeconomic climate indicator (mac\_c), the real-estate climate index (real\_c) and PPI (ppi), and globe market dynamics index, for instance VIX (vix), to be referred than long-term. It thus advises short-term investors to take full considerations, not merely on the price dynamics and market directly related factors, but recruit external economic information, especially the authority labeled index and globe economy dynamics, as well, to improve their risk resistance before the economy sloping down begins.
    
	\begin{sidewaystable}
	l	\caption{Estimated contribution degree of factor variables for currency market: short-term prediction.}
		\setlength{\tabcolsep}{10pt}
		\small
		\begin{center}
			\begin{tabular}{lrrrrrr}
				\hline
				&Coef.$_{LR}^{a}$&NSR$_{KLR}$&Imp.$^{e}_{NN}$&MDA$_{RF}$&Imp.$_{XGBoost}$&Wgt.$^{e}_{AttnLSTM}$\\
				ssec&9.19(0)&2.375&$--^{c}$&4.105&1.07&1.152\\
				ssec\_r&-37.45(0)&0.8562&$--$&0.4071&1.27&1.512\\
				sez&-3.32(0)&2.189&0.1&1.248&2.05&3.326\\
				sez\_r&73.7(0)&0.9599&$--$&1.010&0.55&3.160\\
				frx&-228.2(0)&\textbf{0.311}&\textbf{0.73}&\textbf{5.743}&2.04&1.962\\
				frx\_r&720.3(0)&$--^{b}$&\textbf{0.95}&\textbf{15.29}&\textbf{65.01}&\textbf{5.380}\\
				frx\_res&-2.97(0)&1.559&0.32&0&$--^{d}$&2.507\\
				dis\_r&531.9(0)&1.197&0.16&1.010&0.04&0.7376\\
				int\_r&227.4(0)&Inf$^{b}$&0.22&0&$--$&0.4926\\
				bal\_g&-528(0)&1.596&$--$&0&$--$&0.849\\
				cci&-165.9(0)&\textbf{0.7493}&0.11&4.037&0.36&\textbf{5.252}\\
				cpi&15.22(0)&$--$&$--$&1.733&$--$&2.930\\
				comd\_g&\textbf{2122.3}(0)&1.097&$--$&0&$--$&1.242\\
				deb\_g&\textbf{1198.2}(0)&0.8869&$--$&1.010&$--$&3.517\\
				dom\_c&-0.141(0)&1.018&0.12&0.6468&$--$&3.496\\
				dom\_g&\textbf{-1613}(0)&1.272&$--$&1.720&$--$&3.507\\
				fix\_a&671.5(0)&2.165&0.45&0&0.07&1.030\\
				gdp\_p&-0.717(0)&1.026&$--$&1.010&$--$&3.514\\
				gdp\_p2&\textbf{2587.5}(0)&Inf&\textbf{0.54}&2.058&$--$&0.5556\\
				gdp\_c&-856.5(0)&2.053&$--$&0&0.97&2.300\\
				im\_ex&100.5(0)&$--$&$--$&0&0.50&1.886\\
				ind\_g&\textbf{-1452}(0)&1.013&\textbf{1.53}&1.075&$--$&1.341\\
				ind\_p&-877.4(0)&\textbf{0.7153}&$--$&2.948&\textbf{9.86}&\textbf{5.410}\\
				ind\_v&-346.2(0)&1.150&0.25&1.010&$--$&2.705\\
				inf\_r&-451(0)&\textbf{0.7809}&$--$&1.436&$--$&1.962\\
				mac\_c&\textbf{1075}(0)&\textbf{0.7619}&0.14&2.457&0.05&3.239\\
				m2\_g&447.4(0)&1.437&0.13&1.229&$--$&2.516\\
				ppi&-150.5(0)&$--$&$--$&\textbf{5.210}&\textbf{7.05}&2.958\\
				real\_c&\textbf{1382}(0)&0.9522&$--$&1.602&0.89&1.645\\
				rpi&-299.4(0)&$--$&$--$&2.291&$--$&2.097\\
				rrr&-620.9(0)&\textbf{0.5799}&0.15&0&$--$&3.462\\
				shibor&-917(0)&0.8523&0.48&1.010&$--$&1.578\\
				tax\_r&0.116(0)&1.035&0.16&2.983&2.84&2.399\\
				vix&210.4(0)&0.8092&$--$&1.248&\textbf{5.76}&\textbf{6.038}\\
				gold&16.67(0)&1.569&$--$&2.107&$--$&3.448\\
				oil&61.84(0)&0.9192&$--$&3.535&0.61&1.879\\
				\hline
			\end{tabular}\label{tab:leading_currency_short}
		\end{center}
		
		\footnotesize{$^{a}$ the stepwise logistic regression is adopted. The p-value is labelled in the bracket.}\\
		\footnotesize{$^{b}$ the noise-signal ratio for some variable is either not available or infinity (Inf) since the divider (of the last time point) value is or extremely approaching to zero.}\\
		\footnotesize{$^{c}$ for neural networks model estimation, the importance value below $0.1\%$ will not be displayed.}\\
		\footnotesize{$^{d}$ variables that are assessed to be utterly useless will not be calculated the importance in XGBoost, and the value will be transformed to percentage of $100\%$}\\
		\footnotesize{$^{e}$ both the importance for neural networks and the attention weight for attention-LSTM have been transformed to percentage, i.e. the each weight value times $100\%$.}
	\end{sidewaystable}
	
	\begin{sidewaystable}
		\caption{Estimated contribution degree of factor variables for currency market: long-term prediction.}
		\setlength{\tabcolsep}{10pt}
		\small
		\begin{center}
			\begin{tabular}{lrrrrrr}
				\hline
				&Coef.$_{LR}$&NSR$_{KLR}$&Imp.$_{NN}$&MDA$_{RF}$&Imp.$_{XGBoost}$&Wgt.$_{AttnLSTM}$\\
				ssec&-0.317(0.59)&0.7238&1.01&3.479&1.10&1.938\\
				ssec\_r&$--$&1.086&$--$&0.6342&$--$&2.798\\
				sez&-0.318(0.43)&0.8685&0.79&1.765&$--$&1.948\\
				sez\_r&$--$&1.194&$--$&1.612&$--$&3.452\\
				frx&$--$&\textbf{0.5429}&\textbf{4.52}&1.556&$--$&\textbf{5.812}\\
				frx\_r&$--$&$--$&1.26&0.8380&$--$&2.648\\
				frx\_res&$--$&0.5664&0.25&2.791&$--$&1.457\\
				dis\_r&$--$&1.086&\textbf{3.38}&\textbf{5.681}&\textbf{15.69}&3.252\\
				int\_r&$--$&1.396&$--$&0&$--$&1.394\\
				bal\_g&$--$&\textbf{0.543}&0.37&\textbf{6.422}&$--$&2.962\\
				cci&$--$&0.8143&1.71&1.742&$--$&3.380\\
				cpi&$--$&$--$&1.35&2.698&\textbf{22.97}&3.095\\
				comd\_g&$--$&0.8531&1.49&2.202&4.13&1.688\\
				deb\_g&\textbf{469.2}(0.16)&6.287&$--$&1.663&$--$&1.965\\
				dom\_c&$--$&0.8821&0.41&\textbf{5.114}&\textbf{20.53}&2.2\\
				dom\_g&$--$&0.8143&$--$&1.391&0.8217&2.028\\
				fix\_a&$--$&1.021&$--$&4.175&0.1466&1.448\\
				gdp\_p&-0.190(0.87)&\textbf{0.5428}&$--$&1.381&$--$&1.475\\
				gdp\_p2&$--$&1.086&2.73&\textbf{6.047}&\textbf{22.41}&2.040\\
				gdp\_c&$--$&2.081&\textbf{4.49}&3.787&$--$&3.946\\
				im\_ex&$--$&1.683&$--$&3.405&$--$&1.424\\
				ind\_g&$--$&\textbf{0.5428}&2.21&2.755&0.1445&\textbf{5.038}\\
				ind\_p&$--$&1.357&$--$&2.472&$--$&4.286\\
				ind\_v&$--$&1.154&$--$&1.767&$--$&2.403\\
				inf\_r&$--$&2.171&$--$&1.737&$--$&1.966\\
				mac\_c&$--$&0.9396&1.64&4.445&4.06&1.651\\
				m2\_g&-32.95(0.31)&6.514&$--$&2.451&$--$&2.295\\
				ppi&-40.29(0.34)&0.9952&$--$&1.483&$--$&2.006\\
				real\_c&$--$&1.670&0.74&1.989&1.17&1.437\\
				rpi&101.6(0.14)&1.469&$--$&1.484&$--$&3.799\\
				rrr&69.82(0.16)&0.9870&1.19&2.2&1.12&1.471\\
				shibor&177.8(0.55)&1.163&$--$&1.340&$--$&3.20\\
				tax\_r&$--$&0.7367&$--$&0&$--$&3.116\\
				vix&$--$&1.493&1.3&4.247&5.72&2.795\\
				gold&$--$&1.706&\textbf{5.18}&1.642&$--$&3.1\\
				oil&$--$&1.425&2.36&1.006&$--$&2.564\\
				\hline
			\end{tabular}\label{tabb:leading_currency_long}
		\end{center}
	\end{sidewaystable}
	
	\begin{sidewaystable}
		\caption{Estimated contribution degree of factor variables for stock market: short-term prediction.}
		\setlength{\tabcolsep}{10pt}
		\small
		\begin{center}
			\begin{tabular}{lrrrrrr}
				\hline
				&Coef.$_{LR}$&NSR$_{KLR}$&Imp.$_{NN}$&MDA$_{RF}$&Imp.$_{XGBoost}$&Wgt.$_{AttnLSTM}$\\
				ssec&-4.439(1E-13)&0.815&2.54&\textbf{19.67}&3.723&2.513\\
				ssec\_r&-86.34(7E-04)&0.9288&$--$&6.601&\textbf{6.348}&\textbf{5.467}\\
				sez&0.8479(2E-07)&0.8982&1.25&\textbf{15.94}&\textbf{6.914}&1.485\\
				sez\_r&-99.11(1E-03)&1.013&$--$&7.343&1.021&2.649\\
				frx&$--$&1.502&\textbf{14.8}&9.588&1.816&\textbf{5.605}\\
				frx\_r&$--$&$--$&$--$&8.254&1.182&1.182\\
				frx\_res&$--$&1.256&1.62&5.405&$--$&0.1031\\
				dis\_r&-229.2(6E-07)&Inf&$--$&2.758&$--$&4.603\\
				int\_r&\textbf{3473}(0.03)&inf&$--$&1.750&$--$&4.588\\
				bal\_g&-799.3(9E-11)&0.6569&0.87&2.540&\textbf{7.752}&\textbf{5.974}\\
				cci&69.73(0.135)&0.7915&1.75&4.776&1.858&1.446\\
				cpi&1418(2E-06)&$--$&1.07&7.420&$--$&3.396\\
				comd\_g&577.5(2E-09)&\textbf{0.4778}&$--$&12.11&0.927&\textbf{5.323}\\
				deb\_g&\textbf{-2131}(2E-14)&\textbf{0.5915}&$--$&1.805&$--$&0.0339\\
				dom\_c&0.0592(8E-08)&0.6995&$--$&5.768&3.769&0.1986\\
				dom\_g&-275.6(2E-11)&\textbf{0.4977}&\textbf{5.88}&1.744&$--$&\textbf{5.460}\\
				fix\_a&131.6(3E-06)&0.8305&$--$&6.866&3.231&0.1257\\
				gdp\_p&1.563(1E-11)&0.6271&$--$&2.307&$--$&0.0525\\
				gdp\_p2&$--$&\textbf{0.5773}&0.96&4.483&2.967&\textbf{5.571}\\
				gdp\_c&-175.6(3E-15)&0.8029&3.66&3.308&2.217&1.651\\
				im\_ex&97.78(2E-15)&0.7117&0.65&10.59&3.354&3.221\\
				ind\_g&\textbf{-2292}(5E-04)&1.085&\textbf{5.78}&3.703&$--$&\textbf{5.026}\\
				ind\_p&-374.4(1E-07)&0.7202&$--$&12.83&4.944&0.9214\\
				ind\_v&-99.84(4E-04)&0.8155&$--$&12.57&4.687&4.116\\
				inf\_r&\textbf{-1865}(1E-14)&0.8162&1.90&2.064&$--$&3.4\\
				mac\_c&444.3(4E-08)&1.718&$--$&6.047&$--$&1.529\\
				m2\_g&326.2(5E-09)&\textbf{0.4629}&$--$&2.577&$--$&0.0388\\
				ppi&-548.2(5E-08)&$--$&\textbf{9.37}&10.42&\textbf{9.090}&\textbf{5.399}\\
				real\_c&-409.3(4E-07)&0.9107&4.08&9.425&\textbf{8.144}&2.65\\
				rpi&\textbf{-1803}(6E-07)&$--$&0.34&7.656&\textbf{8.452}&2.707\\
				rrr&964.6(4E-07)&inf&0.67&3.036&$--$&0.4772\\
				shibor&$--$&1.342&$--$&5.992&$--$&2.445\\
				tax\_r&-0.1786(0.03)&0.7438&$--$&8.948&1.234&0.9698\\
				vix&$--$&1.137&$--$&\textbf{14.46}&$--$&\textbf{5.539}\\
				gold&-3.169(0.03)&1.359&\textbf{5.94}&\textbf{19.43}&\textbf{6.864}&0.593\\
				oil&31.61(0.04)&1.189&$--$&12.76&3.856&3.341\\
				\hline
			\end{tabular}\label{tab:leading_stock_short}
		\end{center}
	\end{sidewaystable}
	
	\begin{sidewaystable}
		\caption{Estimated contribution degree of factor variables for stock market: long-term prediction.}
		\setlength{\tabcolsep}{10pt}
		\small
		\begin{center}
			\begin{tabular}{lrrrrrr}
				\hline
				&Coef.$_{LR}$&NSR$_{KLR}$&Imp.$_{NN}$&MDA$_{RF}$&Imp.$_{XGBoost}$&Wgt.$_{AttnLSTM}$\\
				ssec&$--$&0.5134&\textbf{3.77}&\textbf{8.645}&\textbf{53.88}&\textbf{5.451}\\
				ssec\_r&$--$&0.8956&$--$&0.7281&$--$&4.052\\
				sez&$--$&0.6135&2.03&2.795&$--$&1.282\\
				sez\_r&$--$&1.037&$--$&0.8754&2.033&3.119\\
				frx&\textbf{-1489}(0.17)&\textbf{0.4242}&1.45&2.678&1.236&\textbf{6.282}\\
				frx\_r&$--$&$--$&$--$&1.421&0.4794&2.986\\
				frx\_res&-0.063(0.68)&0.5805&$--$&1.865&$--$&0.9285\\
				dis\_r&$--$&$--$&$--$&1.004&$--$&2.766\\
				int\_r&$--$&1.527&$--$&0&$--$&4.581\\
				bal\_g&$--$&\textbf{0.4243}&1.08&\textbf{7.403}&3.08&\textbf{5.395}\\
				cci&$--$&0.8485&0.25&2.44&$--$&3.174\\
				cpi&$--$&$--$&$--$&2.988&$--$&2.074\\
				comd\_g&61.0(0.66)&14.0&0.74&1.644&$--$&1.516\\
				deb\_g&$--$&1.556&$--$&1.886&$--$&1.779\\
				dom\_c&$--$&1.349&$--$&3.069&0.8384&0.9928\\
				dom\_g&$--$&1.1167&0.68&2.087&$--$&1.602\\
				fix\_a&$--$&Inf&0.37&2.109&$--$&1.840\\
				gdp\_p&$--$&\textbf{0.4242}&$--$&3.196&$--$&0.9738\\
				gdp\_p2&$--$&0.8485&1.73&\textbf{5.368}&$--$&1.689\\
				gdp\_c&$--$&1.591&0.50&1.573&$--$&4.643\\
				im\_ex&$--$&1.157&2.27&0.4827&$--$&4.118\\
				ind\_g&\textbf{-230.2}(0.27)&0.8485&2.13&2.418&$--$&4.709\\
				ind\_p&$--$&1.219&$--$&1.004&$--$&3.339\\
				ind\_v&$--$&9.121&$--$&2.656&0.0525&3.423\\
				inf\_r&$--$&9.30&\textbf{4.97}&\textbf{11.34}&\textbf{31.64}&\textbf{5.038}\\
				mac\_c&$--$&1.303&$--$&1.475&$--$&2.842\\
				m2\_g&-6.76(0.86)&1.468&1.26&2.682&$--$&1.851\\
				ppi&$--$&2.121&1.23&2.916&$--$&3.432\\
				real\_c&$--$&1.612&2.53&2.423&2.734&2.140\\
				rpi&$--$&2.439&0.95&2.177&$--$&1.740\\
				rrr&$--$&$--$&$--$&0.7252&$--$&1.605\\
				shibor&$--$&0.9015&3.44&0.8355&$--$&3.014\\
				tax\_r&$--$&0.5874&$--$&1.424&0.0506&1.062\\
				vix&$--$&1.113&$--$&2.439&$--$&3.258\\
				gold&$--$&0.5833&3.33&2.091&1.496&0.9897\\
				oil&-6.82(0.89)&1.023&0.79&1.286&2.482&1.329\\
				\hline
			\end{tabular}\label{tab:leading_stock_long}
		\end{center}
	\end{sidewaystable}

	\section{Conclusions}\label{sec:conclusion}
	This paper contributes to uniformly reorganize the debating arguments on constructing early warning systems to predict financial crises in a full comprehending way to compare two essential components of crisis identifier and crisis predictive models in a unified frameworks, and discuss the variation of leading factors' linkage to sign the market-specific crisis as the deep-going inspection is demanded by policy makers and practitioners to give adjustment and further diminish risks that potentially drive the economic downturn. 
	
    We experiment a full mixing of hybrid EWS models, including classic technical quantified index and Markov switching ARCH model based crisis classifiers, and traditional econometric and state-of-art machine learning based predictive models, on Chinese currency and stock markets in the past twenty-four years daily and monthly data, and mainly get three-fold results. First and foremost, the argument of crisis being predictable seems not be possible to sink but float, even wander, in the debating pool, as the market endogenous instability is proven to be inferred from sufficient forewarning data sources given the modern blossoming powerful predicting techniques. Second, the crisis variable definition should be diversified for a varied predicting period and the SWARCH classified crisis observations is statistically examined more effective as it distinguishes the price index volatility dynamics into several state levels, such as low and high, or low, medium and high, according to the amplitude of fluctuations, and elastically adjusts the value of threshold to augment the crisis identifying flexibility. Third, the SWARCH with automatically optimized threshold combined deep neural networks models cluster outperform other EWS combinations, more specifically, in the view of predicting power, it show greatest robustness given highest precision and widest precautionary periods, and in the practical perspective, it has been back tested to succeed most in terms of lowering the EWS involved portfolio returns instability, especially for daily exercising. Last but not least, the leading factors' variation, the point has yet been discussed in previous studies but begins to start up in our study, is assignable for market-specific crisis prediction. Besides, the predicting term is investigated to negatively effect the quantity of included leading factors, in other words, shorter the term is, more factors will be drawn as leading indicators to foresee crises, albeit in a somewhat overloaded information disturbance in the real world. 
    
    According to the results, we conclude major implications in the practical application perspective. First of all, the necessity of introducing an effective classifier to the crisis prediction is verified in the study, since such crisis identification technique, for example the SWARCH with dynamically optimized cutoff model, is capable of, not only weakening the impact from imbalanced sample distribution, as \cite{CHATZIS2018353} previously mentioned, but boosting the warning system's predicting power as well. Second, the machine learning model based EWS cluster, though often bears the criticism on over-parameterization and weak interpretability (of inner cell parameters and structural weights), does give compelling performance on crisis predicting effectiveness and lowering investment risk in empirical application. It is though hard to dogmatically assert that machine learning based warning systems generally dominate the traditional models, especially considering there should be a great diversity among different types of financial crises, at least, proportional to the market information symmetry, machine learning, especially deep learning, base early warning system should be designated as long as the data sources are abundantly acquired. Third, there is a unilateral contagious effect across the stock and currency crises, that is the exchange rate dynamics can be referred as a precognition factor to forewarn the stock market crisis, but the reverse risk transmission from stocks to currency is yet revealed by evident proofs. Thus, in final, for decision makers, constructing an early warning system is required to considerately examine whether the classification methodology is competent to fully reflect the market fluctuation according to their each volatility rules, in addition to select the most suitable model in line with different objectives, such as either solely pursuing high predicting accuracy or ensuring the output reliability, and either searching for striking factors to crises or implementing mixed aims. On the other hand, in the investor's view, especially for short-term investors with low risk aversion, the requirements to model producing timely and effective warning signals share a greater priority considering they may undertake higher risk of loss in crises, therefore, referring forewarned results from an ensemble of EWS models probably keep them safer than solely relying on one.

    The concluded remarks in the study do not aim for depriving the privilege on open questioning and further exploring the predicting system generalizations, but to achieve the extensibility on the crisis predictability arguments and the comprehensiveness on the early warning system predicting precision and practical value. This study suggests, among a broad aggregation of crisis forecasting methodologies, the crisis identification that relies on the volatility regime switching frameworks partnering with the predictive models that entrust stylized machine learning techniques (especially deep neurons networks), seem to earn rather more accreditation than traditional means under the high-frequency data preliminaries. Further efforts, based on current inspiring findings, are thus worthy of being paid in following aspects: (i) at the methodological level, the diversity of machine learning methods are encouraged to be explored, especially in arithmetic and structures design; (ii) at the computational level, as the computing capacity is continuously boosting, precisely forecasting the timing of crises or crashing events will no longer be unattainable goal if the real-time computation can be introduced; (iii) at the data level, either cross-section or cross-country is yet discussed but highly valued in research on the crisis recurrences for regional and global economies as long as the database accessing is not restricted.
    
    \bibliography{ews_rev.bib}

	\section*{Appendix I}\label{appndx:reference_summary}
     \textit{The table is being processed and will be pasted later...}

	\newpage
	\section*{Appendix II}\label{appndx:classified_crises}
		\begin{figure}[htpb]
	    \centering
	    \vspace{-0.35cm}
	    \subfigtopskip=2pt 
	    \subfigbottomskip=-1pt 
	    \subfigcapskip=-5pt 
	    \subfigure[Exchange rate and SSEC index dynamics.]{
	    \includegraphics[width=0.5\columnwidth]{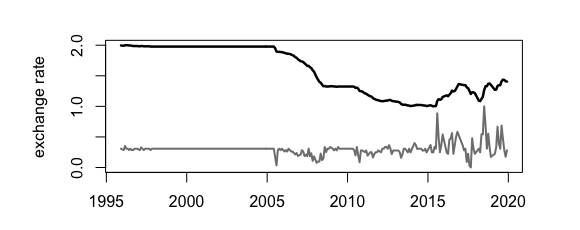}
	    \includegraphics[width=0.5\columnwidth]{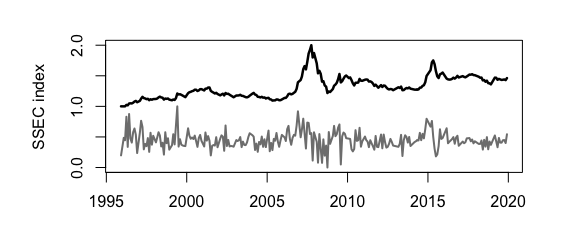}}
	    \subfigure[FPI and CMAX identified crises.] {
      \includegraphics[width=0.5\columnwidth]{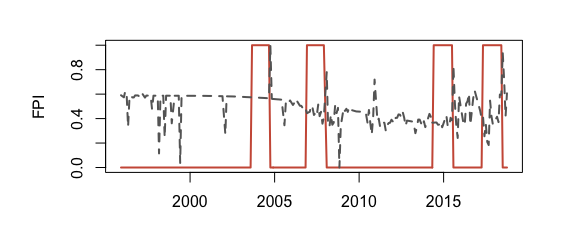} 
      \includegraphics[width=0.5\columnwidth]{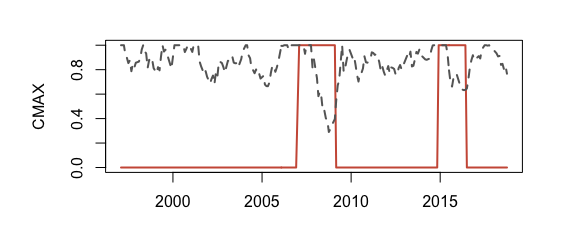}} 
        \subfigure[SWARCH$_{opt}$ identified crises.]{
      \includegraphics[width=0.5\columnwidth]{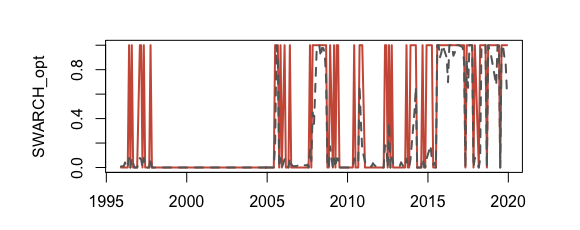}
      \includegraphics[width=0.5\columnwidth]{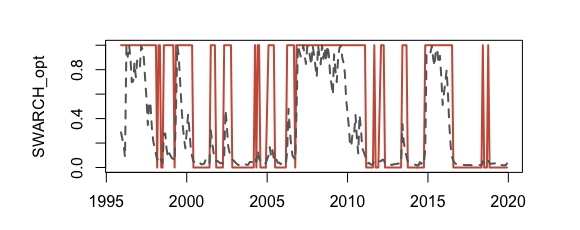}} 
      \caption{Visualized TQI and SWARCH$_{opt}$ identified crises on full samples for monthly data. (a) plots the exchange rate and SSEC index including their change return dynamics (under the price index). The grey dotted lines in (b) label the normalized FPI index and CMAX index and that in (c) label the filtering probability value extracted from AR(1)-SWARCH(2,1). Red solid lines represent the identified crises.} 
       \label{fig:TQI_vs_SWARCH_mth} 
	\end{figure}

\end{document}